\documentclass[12pt]{article}
\usepackage{a4wide,epsfig,psfrag,amsmath,amssymb,cite,scalefnt}
\usepackage{color}
\usepackage{axodraw,slashed}

\newcommand{\mgut}{M_{\rm G}}
\newcommand{\mhc}{M_{\rm H_c}}
\newcommand{\mx}{M_{\rm X}}

\newcommand{\msig}{M_{\rm \Sigma}}

\newcommand{\abbrev}{\scalefont{.9}}
\newcommand{\drbar}{$\overline{\mbox{\abbrev DR}}$}
\newcommand{\msbar}{$\overline{\mbox{\abbrev MS}}$}

\newcommand{\Tt}{\tilde{T}}
\newcommand{\mugut}{\mu_{\rm GUT}}
\newcommand{\one}{\mbox{$1 \hspace{-1.0mm}  {\bf l}$}}
\newcommand{\mbMSbar}{m_{b}^{\overline{\rm MS}}}
\newcommand{\wP}{\varrho}

\begin{document}


\title{\vskip-3cm{\baselineskip14pt
    \begin{flushleft}
      \normalsize SFB/CPP-10-108\\
      \normalsize TTP10-46 
  \end{flushleft}}
  \vskip1.5cm
  Towards a Two-Loop Matching of Gauge Couplings in Grand Unified Theories
}

\author{
  W.~Martens
  \\[1em]
  {\small\it Institut f{\"u}r Theoretische Teilchenphysik}\\
  {\small\it Karlsruhe Institute of Technology (KIT)}\\
  {\small\it 76128 Karlsruhe, Germany}\\
  {\small\it E-Mail: martens@particle.uni-karlsruhe.de}
}

\date{}

\maketitle

\thispagestyle{empty}

\begin{abstract}
  We calculate the two-loop matching corrections for the gauge couplings at the Grand Unification scale
  in a general framework that aims at making as few assumptions on the underlying Grand Unified Theory (GUT)
  as possible.
  In this paper we present an intermediate result that is general enough to
  be applied to the Georgi-Glashow $SU(5)$ as a ``toy model". 
  The numerical effects in this theory are found to be larger than the current experimental uncertainty on $\alpha_s$.
  Furthermore, we give many technical details regarding renormalization procedure,
  tadpole terms, gauge fixing and the treatment of group theory factors, which is useful
  preparative work for the extension of the calculation to supersymmetric GUTs.
\medskip

%
\end{abstract}


\newpage

%
\section{\label{sec::intro}Introduction}
Grand Unified Theories (GUTs) provide an appealing framework for physics beyond 
the Standard Model (SM) of particle physics. Particularly supersymmetric (SUSY) GUTs have gained a lot of
interest in the past decades as they seem to be consistent with the measured values of
$\alpha_s(M_Z)$, $\alpha_{em}(M_Z)$ and $\sin\theta_W(M_Z)$~\cite{Ellis:1990wk,Amaldi:1991cn,Langacker:1991an} 
and offer many other beautiful features.
Up to date most Renormalization Group (RG) analyses that study gauge coupling unification use two-loop 
Renormalization Group Equations (RGEs) and
one-loop matching at the SUSY and 
GUT scales~\cite{Hagiwara:1992ys,Hisano:1992jj,Hisano:1992mh,Yamada:1992kv,
Hisano:1994hb,Dedes:1996wc,Murayama:2001ur,Dorsner:2006ye}. 
These matching corrections
arise from integrating out heavy particles at the SUSY and GUT thresholds. They depend sensitively
on the mass splittings between the heavy particles and can be used to constrain the mass spectrum 
of the theory.
As experimental accuracy is increasing, also higher order corrections become more and more important.
It has been shown that $\mathcal O(\alpha_s^3)$ effects at the SUSY decoupling scale can be as large
as the current experimental uncertainties on the gauge couplings~\cite{Harlander:2005wm,Bauer:2008bj,Martens:2010nm}. 
Furthermore, in GUT models that contain large representations, as for example
the so-called Missing Doublet Model~\cite{Masiero:1982fe,Grinstein:1982um}, the decoupling scale 
dependence at the GUT scale can exceed the experimental
uncertainty by an order of magnitude~\cite{Martens:2010nm}.

These facts encourage us to try to improve the theoretical accuracy of the unification study
and aim at a complete three-loop RGE analysis. This requires three-loop RGEs for the SM, 
the Minimal Supersymmetric Standard Model (MSSM) and the 
SUSY GUT model under consideration and two-loop matching corrections at the SUSY scale and the GUT scale.
In Quantum Chromodynamics (QCD) the gauge $\beta$ function in the modified minimal substraction scheme (\msbar) is known to 
four loops~\cite{vanRitbergen:1997va,Czakon:2004bu}, 
though a complete three-loop SM gauge $\beta$ function is still missing. 
For the MSSM full three-loop RGEs are available~\cite{Harlander:2009mn,Ferreira:1996ug} 
in the so-called~\drbar~scheme. 
The same is true for the most general single gauge coupling theory in \msbar~ and
a general supersymmetric GUT in \drbar~\cite{Pickering:2001aq,Jack:1996vg}.
Matching corrections at the SUSY scale are known to one-loop order for the $U(1)$ and $SU(2)$ 
gauge couplings $\alpha_{1/2}$~\cite{Yamada:1992kv}
and to two-loop for the strong coupling $\alpha_s$~\cite{Harlander:2005wm,Bauer:2008bj}. For the GUT scale thresholds
a general formula at the one-loop level is known~\cite{Hall:1980kf,Weinberg:1980wa,Einhorn:1981sx}, but two-loop
corrections have still been missing up to date and thus are not included in present unification 
analyses~\cite{Hisano:1992mh, Hisano:1992jj, Hagiwara:1992ys, Murayama:2001ur,Dorsner:2006ye,Martens:2010nm}.

The aim of this paper is to provide a first step towards 
a general formula for the two-loop GUT matching corrections in a similar fashion as 
the one-loop corrections in refs.~\cite{Hall:1980kf,Weinberg:1980wa,Einhorn:1981sx}.
Unfortunately it turns out that at the two-loop level it is much harder to carry out the calculation in
a general way, making as few assumptions about the underlying GUT model as possible. 
Therefore, the result given in this paper is not yet 
applicable to SUSY GUT models, as it is not yet general enough.
Nevertheless, by applying it to the Georgi-Glashow $SU(5)$~\cite{Georgi:1974sy} as a ``toy model'', we provide
an important intermediate step on the way to a full three-loop unification analysis. 
The generalization for SUSY GUTs actually is in progress.

\pagebreak

The remainder of the paper is organized as follows: in the next section we describe the theoretical framework
that is used for the calculation. There the Lagrangian is defined and it is shown how delicate issues, as gauge fixing and
renormalization are carried out. In section \ref{sec::num} we present an academic study of our general formula
when applied to the Georgi-Glashow model, which is the simplest (yet already ruled out) possible GUT model. 
In particular the issue of reducing the decoupling scale dependence is discussed. Finally, we present our conclusions. 
In appendix~\ref{sec::group} of this paper we describe the procedure of defining and reducing the group theory factors that 
appear in our calculation. The rest of the appendix is dedicated to some supplementary material to the main text.


\section{\label{sec::theo}Theoretical framework}
Since there is a host of well motivated GUT models, we want our final result to be applicable
to as many of them as possible. Therefore, it is desirable to carry out the calculation of 
two-loop matching corrections at the GUT scale in a framework that makes as few assumptions on the 
underlying GUT model as possible.
The idea is to have a general formula that depends on the Casimir invariants and the spectrum of the 
theory. Choosing a specific model specifies those Casimir invariants and gives an expression that depends 
only on the masses and couplings of the model.
The actual calculation that is presented in this paper is not yet done in full generality, 
but makes some additional assumptions about the model. 
These assumptions are described in subsection \ref{subsec::dec}. 
Nevertheless, we present the theoretical framework that is
needed for the calculation of the relevant Green's functions
(almost) as general as possible below
in order to be armed for future improvements of the calculation.

\subsection{\label{subsec::lag}The Lagrangian}
We consider a general renormalizable quantum field theory defined by the following Lagrangian:
\begin{eqnarray}
\label{eq::gutlag} 
 \mathcal L = -\frac{1}{4}F^{\mu\nu}_\alpha F_{\mu\nu}^\alpha + \overline{\Psi} iD_\mu\gamma^\mu \Psi
+ \frac{1}{2}(D^\mu\Phi)^T D_\mu \Phi - V(\Phi) + \mathcal L_{\rm Y} + \mathcal L_{\rm gf} + \mathcal L_{\rm gh}.
\end{eqnarray}
The chiral Dirac fermion field $\Psi$ and the real scalar field $\Phi$ reside in (not necessarily irreducible)
representations of the gauge group $\mathbf{G}$. The dynamics of the gauge field that transforms according to
the adjoint representation of $\mathbf{G}$, is described by the Yang Mills curl $F^{\mu\nu}_\alpha = 
\partial^\mu A^\nu_\alpha - \partial^\nu A^\mu_\alpha + g f_{\alpha\beta\gamma} A^\mu_\beta A^\nu_\gamma$.
Moreover, $V(\Phi)$, $\mathcal L_{\rm Y}$, $\mathcal L_{\rm gf}$ and $\mathcal L_{\rm gh}$ 
are the scalar potential, the Yukawa interactions, the gauge fixing and the ghost parts of the Lagrangian, respectively. 
They will be described in detail later in this section. $V(\Phi)$ is chosen such that
the scalar field $\Phi=v+\Phi'$ contains one $\mathbf{G}$-irreducible subspace that develops a vacuum expectation value (vev) $v$,
that breaks $\mathbf{G}$ down to the SM gauge group $\prod_k \mathbf{G}_k = SU(3)\times SU(2)\times U(1)$. Models that
have more than one vev of $\mathcal O(\mgut)$, where $\mgut$ is the Grand Unification scale,
are not covered by our framework yet.
The indices $\alpha, \beta, \gamma...$ belong to the adjoint representation and label the generators of $\mathbf{G}$
which fulfill the commutation relations
\begin{equation}\label{eq::commutationrel}
 [T^\alpha, T^\beta] = i f^{\alpha\beta\gamma}\ T^\gamma  \quad  \text{and also:} \quad
 [\tilde{T}^\alpha, \tilde{T}^\beta] = i f^{\alpha\beta\gamma}\ \tilde{T}^\gamma \,,
\end{equation}
with the structure constants $f^{\alpha\beta\gamma}$. We use the tilde to denote
the generators of the real\footnote{This is no loss of generality since every 
complex scalar can be written as two real scalars.}
scalar representation which fulfill $(\Tt^{\alpha})^T = -\Tt^\alpha$ .
The generator that acts on the fermion field satisfies $(T^{\alpha})^\dagger = T^\alpha$.
Again, $T^\alpha$ and $\tilde{T}^\alpha$ need not necessarily be defined
on irreducible representations of $\mathbf{G}$, but can also have block diagonal form. 
In order to distinguish between broken and unbroken generators, we introduce the notation:
\begin{equation}
\label{eq::splitindex} 
 \{\alpha\} = \sum_i\{A_i\} + \sum_i \{a_i\} = \{A\} + \{a\} \,,
\end{equation}
where $A_i$ label the broken generators of $\mathbf{G}$ belonging to the SM-irreducible subspace labeled by $i$.
If there is only one SM-irreducible subspace in the adjoint representation of $\mathbf{G}$,
as e.g. is practically the case in $SU(5)$, we can omit the sub-index $i$.
In contrast, $a_i$ label the unbroken generators belonging to the subgroup\footnote{Please note
the different meanings of the sub-index $i$ when attached to the capital adjoint index opposed
to when attached to a lowercase adjoint index. } $\mathbf{G}_i$:
\begin{eqnarray}\label{eq::genvev}
 \tilde{T}^{a_i} v &=& 0 \,, \nonumber\\
 \tilde{T}^{A_i} v &\neq& 0 \,.
\end{eqnarray}

The Lagrangian in eq.~(\ref{eq::gutlag}) is invariant under local gauge transformations
with the real parameter $\theta = \theta(x)$:
\begin{eqnarray}
 \Psi & \rightarrow & \Psi - i \theta^\alpha T^\alpha\Psi \,, \nonumber\\
 \Phi & \rightarrow & \Phi - i \theta^\alpha \tilde{T}^\alpha\Phi \,, \nonumber\\
 A^\alpha_\mu & \rightarrow & A^\alpha_\mu + f^{\alpha\beta\gamma}
  \theta^\beta A^\gamma_\mu - \frac{1}{g}\partial_\mu\theta^\alpha \,.
\end{eqnarray}
The covariant derivatives are defined as:
\begin{eqnarray}
 D_\mu \Psi &=& (\partial_\mu - i g T^\alpha A^\alpha_\mu)\Psi \,, \nonumber\\
 D_\mu \Phi &=& (\partial_\mu - i g \tilde{T}^\alpha A^\alpha_\mu) \Phi \,.
\end{eqnarray}
Using eq.~(\ref{eq::commutationrel}), eq.~(\ref{eq::genvev}), and
$f^{a_ib_jA_k} =0$ (cf. appendix \ref{sec::group}), the gauge-kinetic term for 
the scalar field $\Phi = v + \Phi'$  can be written as:
\begin{eqnarray}
\label{eq::scalkin} 
\frac{1}{2} (D^\mu\Phi)^T D_\mu\Phi &=& \frac{1}{2}(\partial^\mu\Phi')^T \partial_\mu\Phi'
+ \frac{1}{2}g^2 v \tilde{T}^A \tilde{T}^B v A^\mu_A A_{\mu B} \nonumber\\
&& + ig v \tilde{T}^A \partial_\mu \Phi A^\mu_A 
+ i g^2 f_{ABa} A^\mu_B A_{\mu a} v \tilde{T}^A \Phi' \nonumber\\
&& + ig \Phi' \tilde{T}^\alpha \partial_\mu \Phi' A^\mu_\alpha 
+ \frac{1}{2}g^2 \Phi' \tilde{T}^\alpha \tilde{T}^\beta \Phi' A^\mu_\alpha A_{\mu\alpha} \nonumber\\
&& + g^2 v \tilde{T}^A \tilde{T}^B \Phi' A^\mu_A A_{\mu B} \,,
\end{eqnarray}
where we can identify the (diagonal) gauge boson mass matrix
\begin{equation}
 (\mx)_{A_iB_i} \equiv g^2 v \tilde{T}^{A_i} \tilde{T}^{B_i} v \,,
\end{equation}
with eigenvalues denoted by $M_{{\rm X}_i}$. Again, the sub-index $i$ labels
the SM-irreducible subspace that is meant, because each SM-irreducible subspace
can be assigned to a definite gauge boson mass.
Note that the position of the adjoint indices $\alpha, A_i, a_i...$ is irrelevant.
Furthermore, it is understood that a partial derivative acts only on the single field, which is next to it.
The gauge-kinetic term for the scalars contains the undesired quadratic mixing 
$ig v \tilde{T}^A \partial_\mu \Phi A^\mu_A $ between Goldstone bosons and heavy gauge bosons.
As we will see in a moment, the gauge fixing Lagrangian $\mathcal L_{\rm gf}$ can be chosen in such a way 
that this term is cancelled, at least at tree-level.

In order to fix the gauge, we choose the $R_\xi$ gauge fixing functional~\cite{Fujikawa:1972fe}
\begin{eqnarray}
\label{eq::gaugefix} 
 f_{A_i} &=& \frac{1}{\sqrt{\xi_{1i}}}\partial_\mu A^\mu_{A_i} 
  - i g \sqrt{\xi_{2i}} v \tilde{T}^{A_i} \Phi' \,, \nonumber\\
 f_{a_i} &=& \frac{1}{\sqrt{\eta_i}}\partial_\mu A^\mu_{a_i} \,.
\end{eqnarray}
Note that we have chosen two distinct gauge parameters $\xi_1$ and $\xi_2$ for each SM-irreducible
subspace of the heavy gauge bosons. They renormalize differently and thus can only be equated
with each other after renormalization. Otherwise not all the Green's functions can be made finite.
In the same way each SM group factor receives its own gauge parameter $\eta_i$.
This subtlety arises first at the two-loop level and is not relevant for computing one-loop matching coefficients.
Although there are other ways to treat the gauge fixing~\cite{Hall:1980kf,Santos:1996vt,Girardi:1982by,Gavela:1981ri},
we find this one most convenient for our purposes.
Employing the procedure described e.g. in ref.~\cite{Muta:1998vi},
eq.~(\ref{eq::gaugefix}) gives us the gauge fixing Lagrangian 
and the ghost interactions:
\begin{eqnarray}
\label{eq::gflag} 
 \mathcal L_{\rm gf}^{R_\xi} &=& - \frac{1}{2}\sum_{i} f^2_{A_i} 
     - \frac{1}{2}\sum_{i} f^2_{a_i} \nonumber\\
 &=&\sum_i \left[ -\frac{1}{2 \xi_{1i}} (\partial_\mu A^\mu_{A_i})^2
 - \frac{1}{2}g^2\xi_{2i}\,\Phi' \tilde{T}^{A_i} v v \tilde{T}^{A_i} \Phi' 
 + i g \sqrt{\frac{\xi_{2i}}{\xi_{1i}}} v \tilde{T}^{A_i} \Phi' \partial^\mu A^{A_i}_\mu \right] \nonumber\\ 
 && - \sum_i   \frac{1}{2 \eta_i}(\partial_\mu A^\mu_{a_i})^2 \,,
\end{eqnarray}
\begin{eqnarray}
\label{eq::ghostlag} 
 \mathcal L_{\rm gh}^{R_\xi} &=& \sum_{ij}\Big[ \partial^\mu c_{A_i}^\dagger (\delta_{A_iB_j} \partial_\mu 
 - g f_{A_iB_j\alpha} A^\alpha_\mu) c_{B_j} \nonumber\\ 
 &&- g^2 \sqrt{\xi_{1i}\xi_{2i}}\ v \tilde{T}^{A_i} \tilde{T}^{B_j} v\ c_{A_i}^\dagger c_{B_j}
 - g^2\sqrt{\xi_{1i}\xi_{2i}}\ v \tilde{T}^{A_i} \tilde{T}^{B_j} \Phi' c_{A_i}^\dagger c_{B_j} \Big] \nonumber\\ 
 && - \sum_{ij}\Big[  g (\eta_j/\xi_{1i})^{1/4} f_{A_ib_jB_i}\partial^\mu c_{A_i}^\dagger A^{B_i}_\mu c_{b_j} 
  + i g^2 (\xi_{1i}^{1/4}\eta_j^{1/4}\xi_{2i}^{1/2}) f_{A_ib_jB_i}\ v \tilde{T}^{B_i} 
     \Phi' c_{A_i}^\dagger c_{b_j} \Big] \nonumber\\ 
 && - \sum_{ij}  g\ (\xi_{1j}/\eta_i)^{1/4} f_{a_iB_jA_j}\partial^\mu c_{a_i}^\dagger A^{A_j}_\mu c_{B_j}  \nonumber\\ 
 && + \sum_{i} \partial^\mu c_{a_i}^\dagger (\delta_{a_ib_i}\partial_\mu - g f_{a_ib_ic_i} A^{c_i}_\mu)c_{b_i} \,.
\end{eqnarray}
Here $c_{A_i}$ and $c_{a_i}$ denote the ghost fields belonging to the heavy and light gauge bosons, respectively.
Note again that for an $SU(5)$ GUT we could do the replacement $A_i \rightarrow A$ for the capital adjoint indices
and the notation would become less clumsy. Here, however, we keep the sub-index $i$
in order to stay as general as possible. As mentioned before, after partial integration the term 
$ i g \sqrt{\xi_{2i} / \xi_{1i}}\ v \tilde{T}^{A_i} \Phi' \partial^\mu A^{A_i}_\mu$ in eq.~(\ref{eq::gflag})
exactly cancels the corresponding term in eq.~(\ref{eq::scalkin}) at tree level, where $\xi_{1i} = \xi_{2_i}$ 
is a valid choice. However, when considering higher orders in perturbation theory,
the bare gauge parameters $\xi_{1i}$ and $\xi_{2_i}$
are not equal to each other and the above term must be kept explicitly as a counterterm in our calculation
(cf. also subsection \ref{subsec::ren}).
The quadratic term in eq.~(\ref{eq::gflag}) can be identified with the (unphysical) Goldstone boson mass matrix:
\begin{eqnarray}\label{eq::goldmass}
M^2_{{\rm Gold}} &\equiv& \frac{1}{2}g^2 \sum_i \xi_{2i}\, \tilde{T}^{A_i} v v \tilde{T}^{A_i}  
\end{eqnarray}
with the property ${\rm Tr}(M^2_{{\rm Gold}})=\sum_i \xi_{2i}\, M^2_{{\rm X}_i}\, D_i^{\mathcal A}$,
where $D_i^{\mathcal A}$ is the dimension of the $i$-th SM-irreducible representation
of the heavy gauge bosons.
From the Goldstone theorem it follows that the structure $v\Tt^{A_i}$ projects
on the subspace of Goldstone bosons, i.e. on the subspace that obtains no mass term
from $V(\Phi)$. Hence, the matix $M^2_{{\rm Gold}}$ has non-zero entries
only on the subspace of Goldstone bosons.

For $V(\Phi)$ we consider the most general renormalizable scalar potential with the discrete symmetry
$\Phi\rightarrow -\Phi$:
\begin{equation}
\label{eq::vphi}
 V(\Phi) = -\frac{1}{2}\mu^2_{ij} \Phi_i\Phi_j + \frac{1}{4!} \  \lambda_{ijkl}\ \Phi_i\Phi_j\Phi_k\Phi_l \,,
\end{equation}
with totally symmetric tensors $\mu^2_{ij}$ and $\lambda_{ijkl}$. We impose the requirements
\begin{eqnarray}
\label{eq:lamgaugeinv}
 0 &=& [\mu^2,\tilde{T}^\alpha] \,, \nonumber \\
 0 &=& \Tt^\alpha_{im}\ \lambda_{mjkl} + \Tt^\alpha_{jm}\ \lambda_{imkl} 
     + \Tt^\alpha_{km}\ \lambda_{ijml} + \Tt^\alpha_{lm}\ \lambda_{ijkm} \,,
\end{eqnarray}
in order to make $V(\Phi)$ gauge invariant under $\mathbf{G}$.
The first equation implies that the matrix $\mu^2$ is proportional to the
unit matrix on each subspace irreducible under $\mathbf{G}$.

To break the GUT symmetry, there has to be one
$\mathbf{G}$-irreducible Higgs representation contained in $\Phi_i$ that develops a vev. 
In order to treat the symmetry breaking appropriately, we define the projector $\Pi^\mathcal{H}$ 
on this particular $\mathbf{G}$-irreducible subspace (clearly, $[\Tt^\alpha,\Pi^\mathcal{H}]=0$). 
This subspace is further divided into the subspace of Goldstone bosons and 
the subspace of physical Higgs bosons with projectors $P^\mathcal{G}$
and $P^\mathcal{\tilde{H}}$, respectively
($[\tilde{T}^a,P^\mathcal{G}]=0=[\tilde{T}^a,P^\mathcal{\tilde{H}}]$).
The physical Higgs bosons receive masses
of order $\mgut$ from $V(\Phi)$, the Goldstone bosons do not.
Using the Goldstone boson mass matrix from eq.~(\ref{eq::goldmass}), 
the projector on the space of Goldstone bosons can be explicitly written down as~\cite{Langacker:1980js}:
\begin{eqnarray}\label{eq::PG}
 (P^\mathcal{G})_{ij} = (\Pi^\mathcal{H} - P^\mathcal{\tilde{H}})_{ij} =  
  g^2\tilde{T}^A_{ik}v_k \left(\frac{1}{M^2_X}\right)_{AB} v_l\tilde{T}^B_{lj}  \,.
\end{eqnarray}
Now we can parametrize the scalars in the following way:
\begin{eqnarray}
 \Phi_i = v_i + \Phi'_i = v_i + H_i + G_i + S_i
\end{eqnarray}
where $v$,$\ H$ and $G$ live only on the subspace defined by $\Pi^\mathcal{H}$. 
$S$ parametrizes all the other scalars\footnote{To see that the number of 
Goldstone bosons is equal to the number of broken generators it is also possible to 
define the Goldstone field as  $G^A :=g\left(\frac{i}{\mx}\right)_{AB}v_i\tilde{T}^B_{ij}\Phi'_j$}:
\begin{eqnarray}
(\one-\Pi^\mathcal{H})_{ij}\Phi'_j &=& S_i \,, \nonumber\\
\Pi^\mathcal{H}_{ij}\Phi'_j &=& H_i + G_i \,, \nonumber\\
P^\mathcal{\tilde{H}}_{ij}\Phi'_j &=& H_i \,,\nonumber\\
P^\mathcal{G}_{ij}\Phi'_j &=& G_i \,. 
\end{eqnarray}
First let us focus on the subspace that $\Pi^\mathcal{H}$ projects on. 
In order to develop a vev on this subspace, the parameter $\mu^2_\mathcal{H}$, defined by
$\Pi^\mathcal{H}\mu^2 \equiv \mu^2_\mathcal{H} \one$, has to be positive. 
If this is the case, it is convenient to parametrize
this part of the scalar potential in terms of physical parameters as the Higgs mass,
the heavy gauge boson mass, the gauge coupling $g$ and the tadpole
instead of the unphysical couplings $\mu^2_\mathcal{H}$ and 
\begin{equation}
 \lambda^\mathcal{H}_{ijkl} \equiv 
\lambda_{i'j'k'l'}\Pi^\mathcal{H}_{i'i}\Pi^\mathcal{H}_{j'j}\Pi^\mathcal{H}_{k'k}\Pi^\mathcal{H}_{l'l} \,.
\end{equation}
In principle, this is analogous to what is usually done for the SM Higgs potential~\cite{Passarino}.
Here, however, it is more involved due to the 
appearance of the general invariant tensor $\lambda^\mathcal{H}_{ijkl}$.
Using essentially eq.~(\ref{eq:lamgaugeinv}) and Schur's lemma,
it is possible to rewrite the up to quadratic terms of the potential in 
terms of new parameters $M_H^2$ (diagonal Higgs mass matrix) and $t$ (tadpole). 
For the trilinear and quartic terms this does not seem to be possible at the level of the Lagrangian. 
Thus, for the moment, we leave those terms expressed by the old parameters $\lambda^\mathcal{H}_{ijkl}$. They
have to be eliminated in favor of $M^2_H, g$ and $\mx^2$ at diagram level 
to make our choice of parameters consistent.

Now, including also the scalars on the 
subspace defined by $\one-\Pi^\mathcal{H}$, which is straightforward, the scalar potential can be 
parametrized as follows\footnote{For more details of how this reparametrization is done,
please refer to appendix \ref{sec::reparvphi}.}:
\begin{eqnarray}\label{eq::vphiexp}
 V(\Phi) &=& t \ v_i H_i + \frac{1}{2} (M^2_H)_{ij}H_i H_j + \frac{1}{2}t \ H_i H_i + \frac{1}{2}t \ G_iG_i + \frac{1}{2} (M^2_S)_{ij}S_i S_j  \nonumber\\
   &+& \frac{1}{2} v_i \lambda^\mathcal{H}_{ijkl} H_j G_k G_l + \frac{1}{2} v_i \lambda^\mathcal{H}_{ijkl}H_j H_k G_l +\frac{1}{6}v_i \lambda^\mathcal{H}_{ijkl} H_j H_k H_l \nonumber\\
   &+& \frac{1}{2} v_i \lambda_{ijkl} H_j H_k S_l + v_i \lambda_{ijkl} H_j G_k S_l + \frac{1}{2} v_i \lambda_{ijkl} G_j G_k S_l \nonumber\\
   &+& \frac{1}{2} v_i \lambda_{ijkl} H_j S_k S_l + \frac{1}{2} v_i \lambda_{ijkl} G_j S_k S_l + \frac{1}{6} v_i \lambda_{ijkl} S_j S_k S_l \nonumber\\
   &+& \frac{1}{24} \lambda^\mathcal{H}_{ijkl}  G_i G_j G_k G_l + \frac{1}{6}\lambda^\mathcal{H}_{ijkl} G_i G_j G_k H_l + \frac{1}{6} \lambda^\mathcal{H}_{ijkl} G_i H_j H_k H_l \nonumber\\
   &+& \frac{1}{4} \lambda^\mathcal{H}_{ijkl}  G_i G_j H_k H_l + \frac{1}{24} \lambda^\mathcal{H}_{ijkl} H_i H_j H_k H_l \nonumber\\
   &+& \frac{1}{6} \lambda_{ijkl}  H_i H_j H_k S_l + \frac{1}{2} \lambda_{ijkl}  H_i H_j G_k S_l + \frac{1}{2} \lambda_{ijkl}  H_i G_j G_k S_l  \nonumber\\
   &+& \frac{1}{6} \lambda_{ijkl}  G_i G_j G_k S_l + \frac{1}{4} \lambda_{ijkl}  H_i H_j S_k S_l + \frac{1}{2} \lambda_{ijkl}  H_i G_j S_k S_l  \nonumber\\
   &+& \frac{1}{4} \lambda_{ijkl}  G_i G_j S_k S_l + \frac{1}{6} \lambda_{ijkl}  H_i S_j S_k S_l + \frac{1}{6} \lambda_{ijkl}  G_i S_j S_k S_l   \nonumber\\
   &+& \frac{1}{24} \lambda_{ijkl}  S_i S_j S_k S_l 
\end{eqnarray}
where
\begin{eqnarray}
\label{eq::tad} 
 t &=& -\mu_\mathcal{H}^2 + \frac{1}{6v^2}\ \lambda^\mathcal{H}_{ijkl}v_i v_j v_k v_l \,,\\
\label{eq::higgsmass} 
 (M^2_H)_{ij} &=&  \frac{1}{2} \lambda^\mathcal{H}_{ijkl} v_k v_l 
  - \frac{1}{6 v^2}\ \lambda^\mathcal{H}_{klmn} v_k v_l v_m v_n\ \Pi^\mathcal{H}_{ij} \,,\\
\label{eq::scalmass} 
 (M^2_\mathcal{S})_{ij} &=&  \frac{1}{2} \lambda^\mathcal{S}_{ijkl} v_k v_l - (\mu^2(\one-\Pi^\mathcal{H}))_{ij}\,.
\end{eqnarray}
Due to gauge invariance under the SM group (eqs.~(\ref{eq::genvev}) and (\ref{eq:lamgaugeinv}) ), 
both mass matrices are diagonal and proportional to the unit matrix 
on each SM-irreducible subspace. $M^2_H$ has only non-zero entries on the subspace defined 
by $P^\mathcal{\tilde{H}}$ and $M^2_S$ only on the subspace defined by $(\one-\Pi^\mathcal{H})$ 
(we have defined $\lambda^\mathcal{S}_{ijkl} v_k v_l \equiv (\one-\Pi^\mathcal{H})_{ii'}
(\one-\Pi^\mathcal{H})_{jj'} \lambda^\mathcal{S}_{i'j'kl} v_k v_l$ ). 

It is important to see that $M^2_S$ must have only positive or zero entries. 
If there are negative entries, some of the $S_i$ would develop a vev and our formalism would not apply.
Strictly speaking, we have $(M^2_S)_{ij} < 0$ for the SM Higgs doublet that is contained in $S_i$, 
which would exclude it from our treatment. 
But since in that case the scales involved have the strong hierarchy 
$\mathcal O(M_W)\ll\mathcal O(M_{GUT})$, we can safely set 
the entry to zero here. To do this, some of the $\lambda^\mathcal{S}_{ijkl} v_k v_l$ must be
fine-tuned against the corresponding $(\mu^2(\one-\Pi^\mathcal{H}))_{ij}$ in eq.~(\ref{eq::scalmass})
which is known as the doublet triplet splitting problem, 
inherent to generic GUTs.
\newpage
Note that the classical minimum of the GUT-breaking Higgs potential 
is defined by the equation $t=0$. However, if we compute higher order corrections, 
the parameter $t \equiv 0 - \delta t$, where $\delta t = \mathcal O(\alpha)$ is a counterterm, 
has to be adjusted in such a way that the renormalized Higgs one point function 
is zero at all orders of perturbation theory.

The last term in eq.~(\ref{eq::gutlag}) to be specified is $\mathcal L_{\rm Y} $.
The most general Yukawa interaction of the chiral Dirac fermion multiplet $\Psi$ 
with the real scalar multiplet $\Phi$ can be written as follows:
\begin{eqnarray}
\label{eq::yuklag}
 \mathcal L_{\rm Y} = -\frac{1}{2}\left( Y^k_{ij} \Psi^T_i C \Psi_j 
  \Phi_k + Y^{k\star}_{ij} \Psi^{cT}_i C \Psi^c_j \Phi_k \right)\,.
\end{eqnarray}
Here $Y^k$ is a complex, symmetric matrix and $C \equiv i\gamma_2\gamma_0$ denotes 
the Dirac charge conjugation matrix. $\Psi^c \equiv C \overline{\Psi}^T$ is 
the charge conjugated Dirac spinor (the$\ ^T$ refers only to Lorentz space). 
We take $\Psi = \frac{1}{2}(1-\gamma_5)\Psi$ to be left-handed so that $\Psi^c$ will be right-handed. 
Furthermore, due to gauge invariance $Y^k_{ij}$ satisfies the following relation:
\begin{equation}
\label{eq::yukinv}
 0 = Y^k_{mj} T^{\alpha}_{mi}  + Y^k_{im} T^{\alpha}_{mj}  + Y^m_{ij}\tilde{T}^\alpha_{mk}\,.
\end{equation}
In the following we will need to distinguish between fields with the mass of $\mathcal O(\mgut)$
and massless fields. We will follow the convention of appendix~\ref{sec::group} and use the projectors $P^x_i$ 
for heavy fields and $p^x_i$ for the light fields.

\subsection{\label{subsec::ren}Renormalization}
In order to do a two-loop calculation of the matching corrections, a one-loop renormalization
program has to be carried out for the theory. The counterterms are adjusted in such a way that
all the one-loop Green's functions of the theory are finite. For convenience we use the on-shell scheme
for the mass parameters of the theory and \msbar~for the gauge couplings, the gauge parameters
and the fields. The renormalized Lagrangian is 
obtained from eq.~(\ref{eq::gutlag}) by the following replacements:
{\allowdisplaybreaks  
\begin{eqnarray}\label{eq::renprescr}
 A^{a_i}_\mu \rightarrow \sqrt{Z_{3i}}\ A^{a_i}_\mu\,, \qquad  
  & \qquad A^{A_i}_\mu  \rightarrow \sqrt{Z_{3i}^X}\ A^{A_i}_\mu \,,\nonumber\\
 p^\mathcal{F}_i \Psi \rightarrow \sqrt{Z_{2i}}\ p^\mathcal{F}_i\Psi\,, \qquad  
  & \qquad P^\mathcal{F}_i \Psi \rightarrow \sqrt{Z_{2i}^{h}}\ P^\mathcal{F}_i\Psi \,,\nonumber\\
 c^{a_i} \rightarrow \sqrt{\tilde{Z}_{3i}}\ c^{a_i} \,,\qquad  
   & \qquad c^{A_i} \rightarrow \sqrt{\tilde{Z}_{3i}^X}\ c^{A_i} \,,\nonumber\\
 P^\mathcal{\tilde{H}}_i H \rightarrow \sqrt{Z_{Hi}}\ P^\mathcal{\tilde{H}}_i H \,,\qquad  
   & \qquad  P^\mathcal{G}_i G \rightarrow \sqrt{Z_{Gi}}\  P^\mathcal{G}_i G \,,\nonumber\\
  p^\mathcal{S}_i S \rightarrow \sqrt{Z_{Si}}\  p^\mathcal{S}_i S \,,\qquad  
   & \qquad P^\mathcal{S}_i S \rightarrow \sqrt{Z^{h}_{Si}}\ P^\mathcal{S}_i S \,,\nonumber\\
 M_{{\rm X}_i}^2 \rightarrow Z_{M_{{\rm X}i}}^2\ M_{{\rm X}_i}^2 \,,\qquad  
   & \qquad  M_{H_i}^2 \rightarrow Z_{M_{Hi}}^2\ M_{H_i}^2 \,,\nonumber\\
 M_{F_i} \rightarrow Z_{M_{Fi}}\ M_{F_i} \,,\qquad  
   & \nonumber\\
 \xi_{1i} \rightarrow Z_{\xi_{1i}}\ \xi \,,\qquad  
   & \qquad  \xi_{2i} \rightarrow Z_{\xi_{2i}}\ \xi \,,\nonumber\\
 \eta_{i} \rightarrow Z_{3i}\ \eta \,,  \qquad  
   & \qquad g \rightarrow Z_g\ g \,.
\end{eqnarray}}
Again, we have used the sub-index $i$ to take care of the fact that there might be several SM-irreducible
representations for a field that all renormalize differently. 
No summation is performed over that index. $P^x_i$ and $p^x_i$ are projectors
on  the various SM-irreducible subspaces of heavy and light fields, respectively. 
$M_F$ is the fermion mass matrix
that can arise from the Yukawa interactions eq.~(\ref{eq::yuklag}). 
Presently, heavy fermions are not included in the 
calculation, so the corresponding renormalization constants are only defined for future convenience.

In the following we list the counterterm Feynman rules that are important for our calculation. 
They are obtained by inserting the renormalization prescriptions from eq.~(\ref{eq::renprescr}) 
into eq.~(\ref{eq::gutlag}) and considering the up to quadratic terms.
For each counterterm we give an expression that is valid to arbitrary loop order in the first line
and in the second line a more convenient expression that is valid only for one-loop renormalization.
We use the notation $Z_i \equiv 1-\delta Z_i$ and $t \equiv 0-\delta t$ 
where $\delta Z_i$ and $\delta t$  are of order $\alpha$.
All the parameters that appear in the equations are renormalized ones.

\noindent
Heavy gauge boson: \\
\begin{minipage}{4cm}
  \begin{picture}(50,20)(0,0)
    \Gluon(3,9)(67,9){2}{9}
    \Text(-2,17)[c]{$A\mu$}
    \Text(72,17)[c]{$B\nu$}
    \Text(35,9)[c]{\begin{LARGE}$\times$\end{LARGE}}
    \Text(35,0)[c]{$\longleftarrow$}
    \Text(35,-5)[c]{$k$}
  \end{picture}
\end{minipage}
\begin{minipage}{6.5cm}
 $$= \hspace{0.5cm} i \delta_{AB} \Big[ (Z_3^X - \frac{Z_3^X}{\xi Z_{\xi_1}} + \frac{1}{\xi} - 1) k_\mu k_\nu
   -(Z_3^X - 1)k^2 g_{\mu\nu} + (Z_3^X Z_{\mx}^2 - 1)\mx^2 g_{\mu\nu} \Big] $$
 $$= \hspace{0.5cm} i \delta_{AB} \Big[ -\delta Z_{\xi_1} k_\mu k_\nu
   -(\mx^2 - k^2)\delta Z_3^X g_{\mu\nu} - 2 \delta Z_{\mx} \mx^2 g_{\mu\nu} \Big] $$
\end{minipage}

\noindent
Heavy ghost: \\
\begin{minipage}{4cm}
  \begin{picture}(50,20)(0,0)
    \DashArrowLine(35,9)(3,9){2}
    \DashArrowLine(67,9)(35,9){2}
    \Text(-2,9)[c]{$A$}
    \Text(72,9)[c]{$B$}
    \Text(35,9)[c]{\begin{LARGE}$\times$\end{LARGE}}
    \Text(35,0)[c]{$\longleftarrow$}
    \Text(35,-5)[c]{$k$}
  \end{picture}
\end{minipage}
\begin{minipage}{6.5cm}
$$= \hspace{0.5cm} i \delta_{AB} \Big[ (\tilde{Z}^X_3 - 1) k^2 - (\tilde{Z}^X_3 \sqrt{Z_{\xi_1}} \sqrt{Z_{\xi_2}} Z_{\mx}^2 - 1)\xi \mx^2\Big]$$
$$= \hspace{0.5cm} i \delta_{AB} \Big[ \delta\tilde{Z}^X_3 (\xi \mx^2 - k^2) + (\frac{1}{2}\delta Z_{\xi_1} +\frac{1}{2}\delta Z_{\xi_2} + 2\delta Z_{\mx})\xi \mx^2\Big]$$
\end{minipage}

\noindent
Goldstone boson: \\
\begin{minipage}{4cm}
  \begin{picture}(50,20)(0,0)
    \DashLine(3,9)(67,9){5}
    \Text(-2,9)[c]{$i$}
    \Text(72,9)[c]{$j$}
    \Text(35,9)[c]{\begin{LARGE}$\times$\end{LARGE}}
    \Text(35,0)[c]{$\longleftarrow$}
    \Text(35,-5)[c]{$k$}
  \end{picture}
\end{minipage}
\begin{minipage}{6.5cm}
 $$= \hspace{0.5cm} i P^\mathcal{G}_{ij} \Big[ (Z_G - 1) k^2 - (Z_G Z_{\xi_2} Z_{\mx}^2 - 1)\xi \mx^2\ -\ t\Big]$$
 $$= \hspace{0.5cm} i P^\mathcal{G}_{ij} \Big[ \delta Z_G (\xi \mx^2 - k^2) + (\delta Z_{\xi_2} + 2\delta Z_{\mx})\xi \mx^2\ +\ \delta t\Big]$$
\end{minipage}

\noindent
Physical Higgs boson: \\
\begin{minipage}{4cm}
  \begin{picture}(50,20)(0,0)
    \DashLine(3,9)(67,9){5}
    \Text(-2,9)[c]{$i$}
    \Text(72,9)[c]{$j$}
    \Text(35,9)[c]{\begin{LARGE}$\times$\end{LARGE}}
    \Text(35,0)[c]{$\longleftarrow$}
    \Text(35,-5)[c]{$k$}
  \end{picture}
\end{minipage}
\begin{minipage}{6.5cm}
 $$= \hspace{0.5cm} i \delta_{ij} \Big[ (Z_H - 1) k^2 - (Z_H Z_{M_H}^2 - 1)M_H^2\ -\ t\Big]$$
 $$= \hspace{0.5cm} i \delta_{ij} \Big[ \delta Z_H (M_H^2 - k^2) + 2 \delta Z_{M_H} M_H^2\ +\ \delta t\Big]$$
\end{minipage}

\pagebreak
\noindent
Mixed counterterm: \\
\begin{minipage}{4cm}
  \begin{picture}(50,20)(0,0)
    \DashLine(3,9)(35,9){5}
    \Gluon(35,9)(67,9){2}{4}
    \Text(-2,9)[c]{$i$}
    \Text(77,9)[c]{$A, \mu$}
    \Text(35,9)[c]{\begin{LARGE}$\times$\end{LARGE}}
    \Text(35,0)[c]{$\longleftarrow$}
    \Text(35,-5)[c]{$k$}
  \end{picture}
\end{minipage}
\begin{minipage}{6.5cm}
$$= \hspace{0.5cm} i g \Big[\sqrt{\frac{\xi_2}{\xi_1}}-1\Big] v_k\tilde{T}^A_{ki} k_\mu\nonumber $$
$$= \hspace{0.5cm} i g\ \frac{1}{2}(\delta Z_{\xi_1} - \delta Z_{\xi_2})\ v_k\tilde{T}^A_{ki} k_\mu\nonumber $$
\end{minipage}

\noindent
Heavy fermion: \\
\begin{minipage}{4cm}
  \begin{picture}(50,20)(0,0)
    \ArrowLine(35,9)(3,9)
    \ArrowLine(67,9)(35,9)
    \Text(-2,9)[c]{$i$}
    \Text(72,9)[c]{$j$}
    \Text(35,9)[c]{\begin{LARGE}$\times$\end{LARGE}}
    \Text(35,0)[c]{$\longleftarrow$}
    \Text(35,-5)[c]{$k$}
  \end{picture}
\end{minipage}
\begin{minipage}{6.5cm}
 $$= \hspace{0.5cm} i \delta_{ij} \Big[ (Z^{h}_2 - 1)\slashed{k} - (Z^{h}_2 Z_{M_F} - 1)M_F \Big]P_L$$
 $$= \hspace{0.5cm} i \delta_{ij} \Big[ \delta Z^{h}_2 (M_F - \slashed{k}) + \delta Z_{M_F} M_F \Big]P_L$$
\end{minipage}

\noindent
Higgs tadpole: \\
\begin{minipage}{4cm}
  \begin{picture}(50,20)(0,0)
    \DashLine(3,9)(67,9){5}
    \Text(-2,9)[c]{$i$}
    \Text(67,9)[c]{\begin{LARGE}$\times$\end{LARGE}}
  \end{picture}
\end{minipage}
\begin{minipage}{6.5cm}
 $$= \hspace{0.5cm} i v_i \delta t$$
\end{minipage}

In order to avoid clutter with the notation, we have omitted the sub-index $i$ here. From the context
it is always unambiguous that the SM-irreducible representation of the field under consideration
is meant. 
For our calculation we need the renormalization constants of all mass and gauge parameters at one-loop.
As can be seen from the above Feynman rules, the set of equations that is used to determine
$\delta Z_{\xi_1}$, $\delta Z_{\xi_2}$ and $\delta t$ is overconstrained (cf. also refs.~\cite{Bohm:1986rj,Chankowski:1992er}). 
This provides a useful check for our calculation: 
we computed $\delta Z_{\xi_1}-\delta Z_{\xi_2}$ from the pole of a combination of the heavy gauge boson
and heavy ghost propagator as well as from the mixed Goldstone boson heavy gauge boson propagator.
Both calculations lead to the same result. In the same way $\delta t$ was computed from the pole of the
physical Higgs tadpole as well as from the Goldstone propagator yielding the same result.
Furthermore, the \msbar~ renormalization constant for the gauge coupling $Z_g$ will be
needed to two-loop order (cf. subsection~\ref{subsec::dec}), including
corrections from the Yukawa couplings. We did the calculation in our framework and found agreement
with the known result in the literature~\cite{Machacek:1983tz,Machacek:1983fi,Pickering:2001aq}.

\subsection{\label{subsec::dec}Decoupling of heavy particles}
When studying gauge coupling unification, it is most convenient 
to use RGEs defined in mass-independent renormalization
schemes, such as \msbar~or \drbar. In these ``unphysical'' schemes the computation of $\beta$ functions
is simplified significantly, which makes them most suitable for this application. However, these schemes
have the well known drawback that the decoupling theorem~\cite{Appelquist:1974tg} does not hold in its naive
form. As a consequence observables of low energy processes will 
depend logarithmically\footnote{At the two-loop order also more complicated functions of the 
heavy masses can appear.} on all the heavy particle
masses of the GUT. This is unacceptable, since it would spoil perturbation theory by the presence of large
logarithms $\ln(\mgut/\mu)$, where $\mu$ is the typical energy scale of the process.
The way out of this dilemma is to use an effective 
theory~\cite{Weinberg:1980wa,Ovrut:1980bk,Ovrut:1981sg,Ovrut:1979pk,Ovrut:1980eq,Ovrut:1980dg,Ovrut:1980uv,Ovrut:1981ue,Ovrut:1981bg,Georgi:1994qn},
where the heavy particles are integrated out at the GUT scale. 
This means that the dynamical degrees of freedom of the heavy particles
are removed, which manifestly leads to power-suppressed contributions of $\mathcal O(1/\mgut)$ in the effective
Lagrangian.
Moreover, the effects of the heavy particles are encoded in a multiplicative redefinition of all the masses and 
couplings of the theory. For the case of the gauge coupling this so-called decoupling relation reads:
\begin{eqnarray}
\label{eq::decrelation}
 \alpha_i(\mugut) &=& \zeta_{\alpha_i}(\mugut,\alpha(\mugut),
  M_{\rm h})\ \alpha(\mugut)\,,\nonumber\\
 \alpha &\equiv& \frac{g^2}{4 \pi}\,, \qquad \alpha_i \equiv \frac{g_i^2}{4\pi}\,, \quad i=1,2,3\,.
\end{eqnarray}
Here $\alpha_i$ and $\alpha$ stands for the \msbar~
gauge coupling\footnote{For simplicity we will only speak about 
\msbar~parameters from now on. If a SUSY GUT is considered, 
all \msbar~parameters will be replaced
by \drbar~parameters.} in the effective theory (the SM or the MSSM) and 
full theory (GUT), respectively. 
$\mugut$ is the unphysical scale, at which the decoupling is performed.
At sufficiently high loop order predictions of physical observables must not depend on $\mugut$ anymore.
The remaining dependence on this scale gives us an estimation of the theory uncertainty of the prediction.
$\zeta_{\alpha_i}$ is the so-called matching coefficient that depends on all the mass parameters of the particles
that have been integrated out. They are abbreviated by $M_{\rm h}$ in eq.~(\ref{eq::decrelation}).
The construction of the effective Lagrangian is described in detail e.g. in
refs.~\cite{Steinhauser:2002rq,Chetyrkin:1997un} for the case of QCD. 
It is applicable to our setup without modifications.
Here we only list the formulas that are relevant for the computation of $\zeta_{\alpha_i}$.
For our calculation we find it most convenient to use the 
three-point Green's function with light ghosts and light gauge bosons
as external particles. Applying Slavnov-Taylor identities to this vertex,
the formula for the \msbar~matching coefficient reads:
\begin{equation}\label{eq::zetaalpha}
 \zeta_{\alpha_i} = \left(\frac{Z_g}{Z_{g_i}}
\frac{\tilde{\zeta}^0_{1i}}{\tilde{\zeta}^0_{3i} \sqrt{\zeta^0_{3i}}}\right)^2\,, \quad i=1,2,3\,.
\end{equation}
The \msbar~renormalization constants for the gauge coupling in the full and effective theory are denoted
by $Z_g$ and $Z_{g_i}$, respectively. The bare matching coefficients for the light ghost-gauge-boson vertex, the
light ghost field and the light gauge field respectively are given by:
\begin{eqnarray}
 \tilde{\zeta}^0_{1i} &=& 1 + \Gamma^{0,h}_{Ac^\dagger c, i}(0,0) \,, \nonumber\\
 \tilde{\zeta}^0_{3i} &=& 1 + \Pi^{0,h}_{c,i}(0) \,, \nonumber\\
 \zeta^0_{3i} &=&  1 + \Pi^{0,h}_{A,i}(0)
\end{eqnarray}
i.e. they are computed from the ``hard part''\footnote{This denotes all the diagrams that contain at least
one heavy particle.} of the one particle irreducible Green's functions with zero external momentum.
All the masses of SM particles are set to zero and therefore only masses of $\mathcal O(\mgut)$
appear in the diagrams.
The external particles for these Green's functions are: two light ghosts with one light gauge boson,
two light ghosts and two light gauge bosons, respectively. In all cases the relevant group structure
has to be projected out and in the case of the light gauge boson propagator only the transverse part
is needed.
The index $i$ takes care of the fact that the SM gauge group is not simple and labels whether
an external gauge bosons and ghosts belonging to $U(1)$, $SU(2)$ or $SU(3)$ have to be taken.

At this point a remark concerning the $\mu_{\rm GUT}$ dependence in different
renormalization schemes might be helpful. In mass-dependent renormalization schemes
the threshold effect around the GUT scale is obtained correctly by including
the effect of the heavy GUT particles in the $\beta$ function. As there is no
explicit matching performed in such schemes,
there is no dependence on the unphysical scale $\mu_{\rm GUT}$ at all.
Therefore, the question might arise
how the $\mu_{\rm GUT}$ dependence in mass-independent renormalization schemes,
as in eq.~(\ref{eq::zetaalpha}), comes about.
The answer is that by naively changing the renormalization scheme from
mass-dependent to mass-independent, low energy observables obtain a logarithmic
dependence on an unphysical renormalization scale $\mu$ that has not been present before.
These are the large logarithms $\log\frac{\mu}{\mgut}$ that would spoil perturbation theory
in the naive mass-independent renormalization scheme. In our (effective theory) approach, however,
we absorb these logarithms in a redefinition of the gauge coupling and choose the unphysical
scale $\mu=\mu_{\rm GUT}$ in the vicinity of the heavy GUT masses, which introduces
an explicit dependence of $\zeta_{\alpha_i}$ on $\mu_{\rm GUT}$.

Before we come to the technical details of the calculation, it is in order to describe all the
assumptions that have been made about the underlying GUT model:
\begin{itemize}
 \item There is no trilinear scalar coupling in $V(\Phi)$.
 \item There is only one vev of $\mathcal O(\mgut)$ in the theory.
 \item There are no heavy fermions in the theory.
 \item The heavy gauge bosons decompose in SM-irreducible representations with a common mass.
 \item The GUT-breaking Higgs decomposes into three SM-irreducible representations ($+$ Goldstone bosons) at most.
       They can all have different masses.
 \item The other scalars in the theory decompose in SM-irreducible representations that have
        a common mass ($+$ light scalars).
 \item The light particles in the theory can decompose in arbitrarily many SM-irreducible representations.
\end{itemize}
As can be seen, the main limitation comes from the number of heavy degrees of freedom in the theory.
The above constraints are designed such that the resulting formula for $\zeta_{\alpha_i}$
is applicable to the simplest GUT, the Georgi-Glashow model, yet keeping the calculation as
simple as possible.
The computational framework for our calculation is set up in such a way that it can
be generalized to more heavy degrees of freedom, in order to apply it to SUSY GUTs in the future.

Given the large number of Feynman diagrams, an automated computation is indispensable. The diagrams
were generated with {\tt QGRAF}~\cite{Nogueira:1991ex} and further processed with  
{\tt q2e} and {\tt exp}~\cite{Seidensticker:1999bb,Harlander:1997zb}. 
In the next step we used a {\tt FORM}~\cite{Vermaseren:2000nd} 
implementation of the two-loop topologies of ref.~\cite{Davydychev:1992mt} by the authors
of ref.~\cite{Bauer:2008bj} and also the {\tt FORM} packages  {\tt MINCER}~\cite{Larin:1991fz} and 
{\tt MATAD}~\cite{Steinhauser:2000ry}. Let us emphasize that no assumptions about the mass hierarchies
of the heavy particles have been made. Therefore, the result is valid for arbitrary numerical values
of the mass parameters as long as their mass splitting is not to large which would lead to power
enhanced contributions and spoil perturbation theory.
The reduction of the group theory factors, which was the most time consuming part,
has been automated and implemented in {\tt FORM}.
The relevant group theory and notational conventions are collected in appendix~\ref{sec::group} of this paper.

\begin{figure}[t]
  \centering
  \SetScale{0.5}
%
\begin{minipage}[t]{3.8cm}
\begin{picture}(170,70)(0,0)
\SetColor{Blue}
\SetWidth{3}
\DashCArc(150,50)(40,100,180){8}
\SetColor{Blue}
\SetWidth{3}
\DashCArc(150,50)(40,180,360){8}
\SetColor{Blue}
\SetWidth{3}
\GlueArc(169,72)(30,148,317){4}{6}
\SetColor{Blue}
\SetWidth{3}
\DashCArc(161,62)(32,-22,125){8}
\SetColor{Black} 
\SetWidth{0.5}
\Gluon(60,50)(110,50){4}{4} 
\SetColor{Black}
\SetWidth{0.5}
\Gluon(190,50)(240,50){4}{4} 
\SetColor{Black}
\Vertex(190,50){2}
\Vertex(110,50){2}
\Vertex(144,89){2}
\end{picture}
\end{minipage}
%
\begin{minipage}[t]{3.8cm}
\begin{picture}(170,70)(0,0)
\SetColor{Red}
\SetWidth{3}
\DashCArc(150,90)(40,0,180){16}
\SetColor{Red}
\SetWidth{3}
\DashLine(110,90)(190,90){16}
\SetColor{Blue}
\SetWidth{3}
\GlueArc(150,90)(40,180,270){4}{5}
\SetColor{Blue}
\SetWidth{3}
\GlueArc(150,90)(40,270,360){4}{5}
\SetColor{Black} 
\SetWidth{0.5}
\Gluon(60,50)(150,50){4}{7} 
\SetColor{Black}
\SetWidth{0.5}
\Gluon(150,50)(240,50){4}{7} 
\SetColor{Black}
\Vertex(190,90){2}
\Vertex(110,90){2}
\Vertex(150,50){2}
\end{picture}
\end{minipage}
%
\begin{minipage}[t]{3.8cm}
\begin{picture}(170,70)(0,0)
\SetColor{Blue}
\SetWidth{3}
\Gluon(110,50)(190,50){4}{6}
\SetColor{Blue}
\SetWidth{3}
\GlueArc(150,50)(40,0,180){4}{9}
\SetColor{Black}
\SetWidth{0.5}
\GlueArc(150,50)(40,180,360){4}{9}
\SetColor{Black} 
\SetWidth{0.5}
\Gluon(60,50)(110,50){4}{4} 
\SetColor{Black}
\SetWidth{0.5}
\Gluon(190,50)(240,50){4}{4} 
\SetColor{Black}
\Vertex(110,50){2}
\Vertex(190,50){2}
\end{picture}
\end{minipage}
%
\begin{minipage}[t]{3.8cm}
\begin{picture}(170,70)(0,0)
\SetColor{Black}
\SetWidth{0.5}
\ArrowLine(150,10)(150,90)
\SetColor{Black}
\SetWidth{0.5}
\ArrowArc(150,50)(40,90,180)
\SetColor{Black}
\SetWidth{0.5}
\ArrowArc(150,50)(40,180,270)
\SetColor{Blue}
\SetWidth{3}
\GlueArc(150,50)(40,0,90){4}{4}
\SetColor{Blue}
\SetWidth{3}
\GlueArc(150,50)(40,270,360){4}{4}
\SetColor{Black} 
\SetWidth{0.5}
\Gluon(60,50)(110,50){4}{4} 
\SetColor{Black}
\SetWidth{0.5}
\Gluon(190,50)(240,50){4}{4} 
\SetColor{Black}
\Vertex(190,50){2}
\Vertex(110,50){2}
\Vertex(150,10){2}
\Vertex(150,90){2}
\end{picture}
\end{minipage}
%

%
%
\begin{minipage}[b]{3.8cm}
\begin{picture}(170,90)(0,0)
\SetColor{Blue}
\SetWidth{3}
\DashArrowArc(150,50)(40,180,360){2}
\SetColor{Blue}
\SetWidth{3}
\DashCArc(150,50)(40,130,180){8}
\SetColor{Blue}
\SetWidth{3}
\GlueArc(150,50)(40,0,50){4}{3} 
\SetColor{Blue}
\SetWidth{3}
\GlueArc(150,82)(23,180,360){4}{6}
\SetColor{Blue}
\SetWidth{3}
\DashCArc(150,82)(23,0,180){8}
\SetColor{Black} 
\SetWidth{0.5}
\DashArrowLine(60,50)(110,50){2} 
\SetColor{Black}
\SetWidth{0.5}
\DashArrowLine(190,50)(240,50){2} 
\SetColor{Black}
\Vertex(190,50){2}
\Vertex(110,50){2}
\Vertex(173,81){2}
\Vertex(127,81){2}
\end{picture}
\end{minipage}
%
%
\begin{minipage}[b]{3.8cm}
\begin{picture}(170,90)(0,0)
\SetColor{Blue}
\SetWidth{3}
\DashArrowArc(150,50)(40,180,360){2}
\SetColor{Blue}
\SetWidth{3}
\GlueArc(150,50)(40,90,180){4}{4}
\SetColor{Blue}
\SetWidth{3}
\GlueArc(150,50)(40,0,90){4}{4} 
\SetColor{Green}
\SetWidth{3}
\DashCArc(150,115)(25,-90,270){8}
\SetColor{Black} 
\SetWidth{0.5}
\DashArrowLine(60,50)(110,50){2} 
\SetColor{Black}
\SetWidth{0.5}
\DashArrowLine(190,50)(240,50){2} 
\SetColor{Black}
\Vertex(190,50){2}
\Vertex(110,50){2}
\Vertex(150,90){2}
\end{picture}
\end{minipage}
%
%
\begin{minipage}[b]{3.8cm}
\begin{picture}(170,90)(0,0)
\SetColor{Blue}
\SetWidth{3}
\DashArrowArc(150,50)(40,180,360){2}
\SetColor{Blue}
\SetWidth{3}
\GlueArc(150,50)(40,130,180){4}{3}
\SetColor{Blue}
\SetWidth{3}
\GlueArc(150,50)(40,0,50){4}{3} 
\SetColor{Black}
\SetWidth{0.5}
\DashCArc(150,82)(23,180,360){8}
\SetColor{Green}
\SetWidth{3}
\DashCArc(150,82)(23,0,180){8}
\SetColor{Black} 
\SetWidth{0.5}
\DashArrowLine(60,50)(110,50){2} 
\SetColor{Black}
\SetWidth{0.5}
\DashArrowLine(190,50)(240,50){2} 
\SetColor{Black}
\Vertex(190,50){2}
\Vertex(110,50){2}
\Vertex(173,81){2}
\Vertex(127,81){2}
\end{picture}
\end{minipage}
%
%
\begin{minipage}[b]{3.8cm}
\begin{picture}(170,90)(0,0)
\SetColor{Blue}
\SetWidth{3}
\Gluon(150,90)(150,10){4}{6}
\SetColor{Blue}
\SetWidth{3}
\DashArrowArc(150,50)(40,270,360){2}
\SetColor{Black}
\SetWidth{0.5}
\DashArrowArc(150,50)(40,180,270){2}
\SetColor{Blue}
\SetWidth{3}
\GlueArc(150,50)(40,0,90){4}{4}
\SetColor{Black}
\SetWidth{0.5}
\GlueArc(150,50)(40,90,180 ){4}{4}
\SetColor{Black} 
\SetWidth{0.5}
\DashArrowLine(60,50)(110,50){2} 
\SetColor{Black}
\SetWidth{0.5}
\DashArrowLine(190,50)(240,50){2} 
\SetColor{Black}
\Vertex(190,50){2}
\Vertex(110,50){2}
\Vertex(150,10){2}
\Vertex(150,90){2}
\end{picture}
\end{minipage}
%
%

%
\begin{minipage}[b]{3.8cm}
\begin{picture}(170,100)(0,0)
\SetColor{Blue}
\SetWidth{3}
\Gluon(190,50)(150,115){4}{6}
\SetColor{Blue}
\SetWidth{3}
\DashArrowLine(110,50)(190,50){2}
\SetColor{Blue}
\SetWidth{0.5}
\DashLine(150,115)(139,96){8}
\SetColor{Blue}
\SetWidth{3}
\Gluon(123,70)(110,50){4}{3}
\SetColor{Green}
\SetWidth{3}
\DashCArc(131,83)(15,55,241){8}
\SetColor{Black}
\SetWidth{0.5}
\DashCArc(131,83)(15,-119,55){8}
\SetColor{Black}
\SetWidth{0.5}
\DashArrowLine(60,50)(110,50){2} 
\SetColor{Black}
\SetWidth{0.5}
\Gluon(150,115)(150,150){4}{3}
\SetColor{Black}
\SetWidth{0.5}
\DashArrowLine(190,50)(240,50){2} 
\SetColor{Black}
\Vertex(110,50){2}
\Vertex(139,96){2}
\Vertex(123,70){2}
\Vertex(150,115){2}
\Vertex(190,50){2}
\end{picture}
\end{minipage}
%
\begin{minipage}[b]{3.8cm}
\begin{picture}(170,100)(0,0)
\SetColor{Blue}
\SetWidth{3}
\DashLine(110,50)(127,77){8}
\SetColor{Blue}
\SetWidth{3}
\Gluon(127,77)(150,115){4}{3}
\SetColor{Blue}
\SetWidth{3}
\DashArrowLine(110,50)(190,50){2}
\SetColor{Blue}
\SetWidth{3}
\Gluon(150,115)(173,77){4}{3}
\SetColor{Blue}
\SetWidth{3}
\Gluon(173,77)(190,50){4}{3}
\SetColor{Red}
\SetWidth{3}
\DashCArc(150,115)(45,239,301){16}
\SetColor{Black}
\SetWidth{0.5}
\DashArrowLine(60,50)(110,50){2} 
\SetColor{Black}
\SetWidth{0.5}
\Gluon(150,115)(150,150){4}{3}
\SetColor{Black}
\SetWidth{0.5}
\DashArrowLine(190,50)(240,50){2} 
\SetColor{Black}
\Vertex(127,77){2}
\Vertex(173,77){2}
\Vertex(150,115){2}
\Vertex(110,50){2}
\Vertex(190,50){2}
\end{picture}
\end{minipage}
%
\begin{minipage}[b]{3.8cm}
\begin{picture}(170,100)(0,0)
\SetColor{Red}
\SetWidth{3}
\DashLine(110,100)(150,140){16}
\SetColor{Red}
\SetWidth{3}
\DashLine(150,140)(190,100){16}
\SetColor{Blue}
\SetWidth{3}
\Gluon(110,50)(110,100){4}{4}
\SetColor{Blue}
\SetWidth{3}
\DashArrowLine(190,100)(190,50){2}
\SetColor{Blue}
\SetWidth{3}
\DashArrowLine(110,50)(190,100){2}
\SetColor{Blue}
\SetWidth{3}
\Gluon(190,50)(110,100){4}{8}
\SetColor{Black}
\SetWidth{0.5}
\DashArrowLine(60,50)(110,50){2} 
\SetColor{Black}
\SetWidth{0.5}
\Gluon(150,140)(150,180){4}{4}
\SetColor{Black}
\SetWidth{0.5}
\DashArrowLine(190,50)(240,50){2} 
\SetColor{Black}
\Vertex(150,140){2}
\Vertex(110,50){2}
\Vertex(190,50){2}
\Vertex(110,100){2}
\Vertex(190,100){2}
\end{picture}
\end{minipage}
%
\begin{minipage}[b]{3.8cm}
\begin{picture}(170,100)(0,0)
\SetColor{Blue}
\SetWidth{3}
\DashArrowLine(110,50)(150,50){2}
\SetColor{Blue}
\SetWidth{3}
\DashArrowLine(150,50)(190,50){2}
\SetColor{Blue}
\SetWidth{3}
\Gluon(110,50)(150,115){4}{6}
\SetColor{Blue}
\SetWidth{3}
\Gluon(150,115)(190,50){4}{6}
\SetColor{Black}
\SetWidth{0.5}
\Gluon(150,50)(150,115){4}{6}
\SetColor{Black}
\SetWidth{0.5}
\DashArrowLine(60,50)(110,50){2} 
\SetColor{Black}
\SetWidth{0.5}
\Gluon(150,115)(150,150){4}{3}
\SetColor{Black}
\SetWidth{0.5}
\DashArrowLine(190,50)(240,50){2} 
\SetColor{Black}
\Vertex(110,50){2}
\Vertex(150,115){2}
\Vertex(150,50){2}
\Vertex(190,50){2}
\end{picture}
\end{minipage}
  \caption{\label{fig::dias}
      Sample two-loop diagrams that appear in the calculation of $\zeta_{\alpha_i}$.
      The first line shows the process $A^{a_i}_\mu \rightarrow A^{b_i}_\nu$ 
      contributing to $\Pi^{0,h}_{A,i}(0)$. 
      The second and third line depict $c^{a_i} \rightarrow c^{b_i}$ and $c^{a_i} \rightarrow c^{b_i} + A^{c_i}_\mu $
      contributing to $\Pi^{0,h}_{c,i}(0)$ and $\Gamma^{0,h}_{A c^\dagger c,i}(0,0)$, respectively.
      Colored (bold) lines represent fields with mass of $\mathcal O(\mgut)$ and black (thin) 
      lines massless fields.
      Furthermore, curly lines denote gauge bosons, dotted lines ghosts, dashed lines scalar fields and
      solid lines fermions. 
      Goldstone bosons are marked green (light gray, short-dashed), 
      physical Higgs bosons red (gray, long-dashed) and other heavy scalars blue (dark, short-dashed).
      Note also that two identical lines in one diagram need not have the same 
      mass because of the non-degenerate mass spectrum.}
\end{figure}
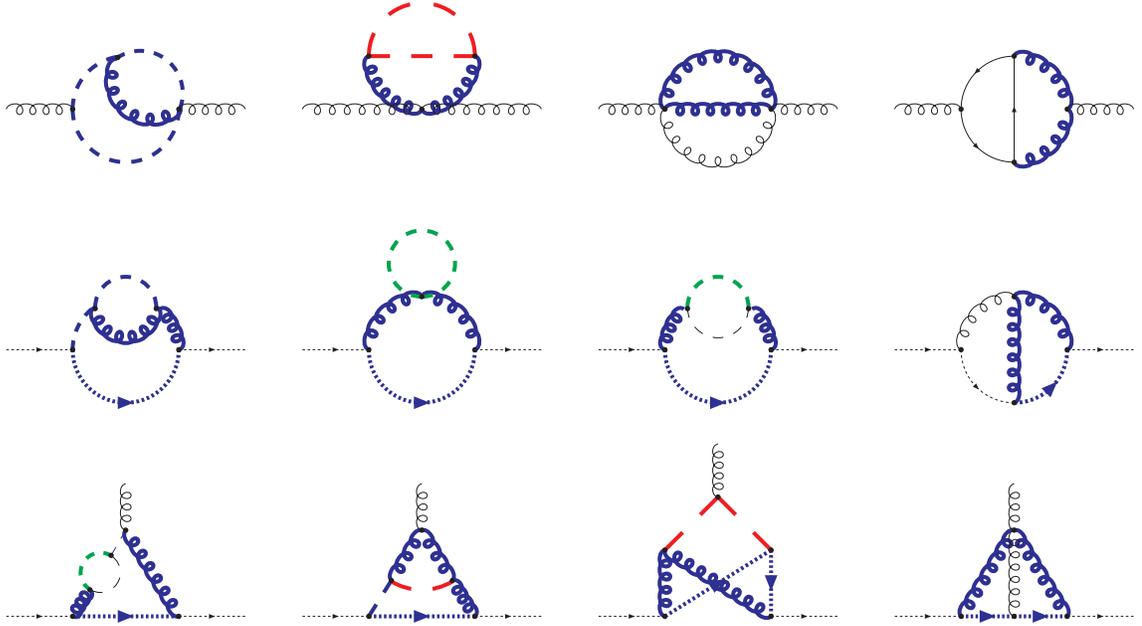

Although the present calculation is just general enough to be applied to the simplest GUT,
the number of Feynman diagrams for the two-loop Green's functions described above already is considerable.
For the light gauge boson two-point function it amounts to 6278, whereas for the ghost-gauge-boson vertex
and the light ghost two-point function we have 4109 and 374 diagrams, respectively.
Sample diagrams\footnote{The figure has been created with help of the \LaTeX~package
{\tt AXODRAW}~\cite{Vermaseren:1994je}} for all three processes are depicted in fig. \ref{fig::dias}.

In order to subtract the subdivergencies that occur in these two-loop Green's functions,
the mass and gauge parameters as well as the gauge coupling
that appear in the corresponding one-loop Green's functions
have to be renormalized as described in subsection \ref{subsec::ren}. We have performed our calculation
for arbitrary gauge parameters and verified that they cancel out in the final result,
which is a powerful and highly non-trivial check of the calculation.
The result for $\zeta_{\alpha_i}$ is available in general form, i.e. 
with group theory factors and couplings not specified to a particular Lie group or model.
In the next section we will assign definite values to these quantities in order
to show the application of the result exemplarily. Note also that we always need three different sets
of group theory factors for $i=1,2,3$, respectively.

\section{\label{sec::num}Numerical study for the Georgi-Glashow model}

In the following we perform an academic study of our results applied to the simplest possible GUT,
the Georgi-Glashow model~\cite{Georgi:1974sy}. Although this theory is ruled out experimentally~\cite{Amaldi:1991cn},
we use it as a toy model to demonstrate some simple numerics. It is based on the gauge group $SU(5)$
and the SM fermions sit in the representations
\begin{eqnarray}
\label{eq::5bar} 
\overline{\mathbf{5}} & \rightarrow & (\overline{\mathbf{3}},\mathbf{1},\tfrac{1}{3}) 
  \oplus (\mathbf{1},\overline{\mathbf{2}},-\tfrac{1}{2}) \,, \\
\label{eq::10} 
\mathbf{10} = \left[\mathbf{5}\times \mathbf{5}\right]_a & \rightarrow & 
 (\overline{\mathbf{3}},\mathbf{1},-\tfrac{2}{3}) \oplus (\mathbf{3},\mathbf{2},\tfrac{1}{6})_a 
 \oplus (\mathbf{1},\mathbf{1},1) \,.
\end{eqnarray} 
where we have displayed their decomposition into SM multiplets.\footnote{The first and second number
label the $SU(3)$ and $SU(2)$ representations, respectively and the third number is the hypercharge of the multiplet.
The index $a$ indicates that the antisymmetric part of the representation has to be taken.}
The gauge bosons as well as the $SU(5)$-breaking Higgs bosons $\Sigma$ 
live in the real $\mathbf{24}$ representation:
\begin{equation}
\mathbf{24} \rightarrow (\mathbf{8},\mathbf{1},0) 
  \oplus (\mathbf{1},\mathbf{3},0) \oplus (\mathbf{1},\mathbf{1},0) 
  \oplus (\mathbf{3},\overline{\mathbf{2}},-\tfrac{5}{6}) 
  \oplus (\overline{\mathbf{3}},\mathbf{2},\tfrac{5}{6}) \,.
\end{equation} 
In the case of the gauge bosons the first three parts of the decomposition constitute the light
SM gauge bosons and the last two parts the heavy gauge bosons with a common mass $\mx$.
For the GUT-breaking Higgs bosons the first three parts represent the three physical Higgs bosons
with masses $\msig$, $2\, \msig$ and $M_{24}$, respectively. The last two multiplets
give the Goldstone bosons with the unphysical mass $\sqrt{\xi_2}\,\mx$ (cf. eq.~(\ref{eq::goldmass})).
Note that we write the field $\Sigma$ as a 24-dimensional vector multiplet
in order to be consistent with our notation in section~\ref{sec::theo}, and not as a
hermitian $5\times 5$ matrix as usually done. Therefore, the vev of $\Sigma$ simply is
$\langle\Sigma\rangle=v$, where $v$ is a 24-dimensional vector with a single non-zero entry
in the 24th component.
Additionally we have a complex scalar in the $\mathbf{5}$ representation that decomposes according
to eq.~(\ref{eq::5bar}) and contains the SM Higgs doublet as well as a heavy colored triplet
with the mass $\mhc$.

For the convenience of the reader it might be helpful to explicitly give the parametrization
of the quartic scalar coupling $\lambda_{ijkl}$ from eq.~(\ref{eq::vphi}) for 
the Georgi-Glashow model. It splits up into
three parts. The first part is a quartic coupling of the $\mathbf{24}$ Higgs, the second one
a quartic coupling of the $\mathbf{5}$ Higgs and the last one is a mixed $\mathbf{5}-\mathbf{24}$ 
coupling~\cite{Sherry:1979sz,Langacker:1980js}:
\begin{eqnarray}
\label{eq::su5lam}
\lambda^{24}_{\alpha \beta \gamma \delta} &=& A\ {\rm sTr}(T^\alpha T^\beta T^\gamma T^\delta) 
 + \frac{1}{3}\ B\ (\delta_{\alpha \beta} \delta_{\gamma \delta} + \delta_{\alpha \gamma} \delta_{\beta \delta} 
 + \delta_{\alpha \delta} \delta_{\beta \gamma} ) \,, \nonumber \\
\lambda^5_{i j k l} &=& \frac{1}{3}\ b\ (\delta_{i j} \delta_{k l} + \delta_{i k} \delta_{j l} 
 + \delta_{i l} \delta_{j k} ) \,, \quad i,j,k,l = 1,...,10 \,, \nonumber \\
\lambda^{5-24}_{\alpha \beta i j} &=& c\ (\tau^\alpha \tau^\beta + \tau^\beta \tau^\alpha)_{i j} \,, \qquad 
  \alpha, \beta, \gamma, \delta = 1,...,24 \,.
\end{eqnarray}
We have used the symmetrized trace
\begin{eqnarray}
 {\rm sTr}(T^{\alpha_1}...T^{\alpha_n}) \equiv \frac{1}{n!} \sum_\pi {\rm Tr}(T^{\alpha_{\pi(1)}}...T^{\alpha_{\pi(n)}})
\end{eqnarray}
where the sum is over all the permutations of the indices. 
Furthermore, it is noticeable that the indices $i,j,k,l$
run from 1 to 10 although they belong to the fundamental $\mathbf{5}$ representation. 
This is because the corresponding scalar is complex and we have written 
it as twice as many real scalars that are transformed by the $10\times 10$ generator
matrices $\tau$:
\begin{equation}
 \tau^\alpha = \left(\begin{array}{cc}
 i\ {\rm Im}(T^\alpha) & i\ {\rm Re}(T^\alpha) \\ 
 -i\ {\rm Re}(T^\alpha) & i\ {\rm Im}(T^\alpha) \\ 
\end{array} \right) \,,
\end{equation}
where in this section $T^\alpha$ is the $5\times 5 $ generator matrix in the fundamental representation.
Inserting eq.~(\ref{eq::su5lam}) into eqs. (\ref{eq::higgsmass}) and (\ref{eq::scalmass}), we obtain the 
scalar mass matrices. Additionally we need to impose the tree-level fine-tuning condition 
$\mu_{\mathbf{5}}^2 = \tfrac{3}{20} c v^2$, where $\mu_{\mathbf{5}}^2$ is the quardatic
term of the $\mathbf{5}$ Higgs in eq.~(\ref{eq::vphi}), in order to obtain massless Higgs doublets.
Note that in principle one would need to calculate the one-loop fine-tuning condition
in order to obtain massless Higgs doublets in a two-loop calculation. However, the light
Higgs doublets show up at the first time in the two-loop Green's functions in the matching calculation 
so that it is sufficient to use the tree-level fine-tuning condition. 
This gives us the relations between the physical scalar masses and the parameters in eq.~(\ref{eq::su5lam}):
\begin{eqnarray}\label{eq::SU5scalarmasses}
 \msig^2 = \tfrac{1}{144}\ A\ v^2 \,, \quad M^2_{24} = \tfrac{1}{3}(\tfrac{7}{120} A + B)\ v^2 \,, \quad
  \mhc^2 = \tfrac{1}{12}\ c\ v^2 \,.
\end{eqnarray}
The vev is connected to the physical gauge boson mass by
\begin{equation}\label{eq::SU5gaugebosonmass}
 \mx^2 = \tfrac{5}{12}\ g^2\ v^2 \,.
\end{equation}

The Yukawa interactions of the Georgi-Glashow model~\cite{Georgi:1974sy,Langacker:1980js} 
are obtained by inserting the Yukawa matrix
\begin{eqnarray}
\label{eq::su5yuk}
  Y^n_{sr} = \left(\begin{array}{cc}
 - Y^U_{IJ}\epsilon_{ijklm}\ T^\alpha_{ij} T^\beta_{kl} S^*_{mn}  &  2 i\ Y^D_{IJ}\  {\rm Im}(T^\alpha_{kl}) S_{ln} \\ 
 2 i\ Y^D_{IJ}\ {\rm Im}(T^\alpha_{kl}) S_{ln}  & 0 \\ 
\end{array} \right)_{sr} 
\end{eqnarray}
into the general Yukawa Lagrangian eq.~(\ref{eq::yuklag}). Here $s=(I,\tilde{s})$ and $r=(J,\tilde{r})$ 
are multi-indices, where $I,J$ stand for the generation indices of the $SU(5)$ Yukawa matrices $Y^U$ and $Y^D$. 
The indices $\tilde{s},\tilde{r} =1,...,29$ run over $\{\alpha, j\}$ and $\{\beta, k \}$, respectively.
Note that we have written the fermions of the $\mathbf{10}$ representation as a $24$-dimensional vector
instead of an antisymmetric $5\times 5$ matrix as usually. The Clebsch-Gordan coefficients for this 
transformation are given by the following equations:
\begin{equation}
 \mathbf{10}^\alpha = \sqrt{2}\ T^\alpha_{ij}\ \mathbf{10}_{ij} \,, \quad
 \mathbf{10}_{ij} =  - \sqrt{2}\ i\ {\rm Im}(T^\alpha_{ij})\ \mathbf{10}^\alpha \,,
\end{equation}
where $\mathbf{10}_{ij}$ is the usual antisymmetric $5\times 5$ matrix with the 
normalization as in ref.~\cite{Georgi:1974sy}. Furthermore,
$S = \frac{1}{\sqrt{2}}(\one,\, i\one)$ is a $5\times 10$ matrix and $\epsilon_{ijklm}$ is the
totally antisymmetric tensor with $\epsilon_{12345}=1$. As can be seen from eq.~(\ref{eq::su5yuk}),
the chiral fermion multiplet $\Psi$ from subsection \ref{subsec::lag} is written as a 
$3\, (24 + 5) = 87$-dimensional vector for the case of Georgi-Glashow $SU(5)$ model.

Using the definitions from this section, we computed the numerical values of all the 
group theory factors that appear in our general result. Furthermore, we set
$V_{\rm CKM} = \one$ and kept only the third generation Yukawa couplings
$y_t$ and $y_b$. We obtained three two-loop
formulas for $\zeta_{\alpha_i}\ (i=1,2,3)$ that depend on the parameters
\begin{equation}
 \alpha(\mugut),\ y_t(\mugut),\ y_b(\mugut),\ \mx,\ \mhc,\ \msig,\ M_{24},\ \mugut \,.
\end{equation}
The  {\tt Mathematica} package that contains the expressions can be downloaded from\\[0.3cm]
{\tt http://www-ttp.particle.uni-karlsruhe.de/Progdata/ttp10/ttp10-46/}

In order to examine the numerical impact of the two-loop matching corrections in this model,
we have implemented a RGE analysis in {\tt Mathematica}. Since in this model the gauge couplings do not unify,
we just focus on examining the reduction of the decoupling scale dependence, as an illustration of our results.
We start with the precise values of the three gauge couplings at the electroweak scale.
They are obtained from the effective weak mixing angle in the $\overline{\rm MS}$
scheme~\cite{Amsler:2008zzb}, the QED coupling constant at zero
momentum transfer and its hadronic~\cite{Teubner:2010ah}
contribution in order to obtain its counterpart at the $Z$-boson scale,
and the strong coupling constant~\cite{Bethke:2009jm}.\footnote{We adopt the central
value from ref.~\cite{Bethke:2009jm}, however, use as our default
choice for the uncertainty $0.0020$ instead of $0.0007$.} 
These quantities need to be transformed to a six-flavor theory, which is described
in detail in ref.~\cite{Martens:2010nm}. Our starting values are then:
\begin{eqnarray}
  \alpha_{em}^{(6),\overline{\rm MS}}(M_Z) &=& 1/(128.129\pm 0.021)
  \,,\nonumber\\
  \sin^2\Theta^{(6),\overline{\rm MS}}(M_Z) &=& 0.23138 \pm 0.00014
  \,,\nonumber\\
  \alpha_s^{(6)}(M_Z) &=& 0.1173\pm 0.0020
  \,.
  \label{eq::alphasin}
\end{eqnarray}
These quantities are related to the three gauge couplings via
\begin{eqnarray}
  \alpha_1 &=& \frac{5}{3}
  \frac{\alpha_{em}^{(6),\overline{\rm MS}}}{\cos^2\Theta^{(6),\overline{\rm MS}}}
  \,,\nonumber\\
  \alpha_2 &=& \frac{\alpha_{em}^{(6),\overline{\rm
        MS}}}{\sin^2\Theta^{(6),\overline{\rm MS}}}
  \,,\nonumber\\
  \alpha_3 &=& \alpha_s^{(6)}
  \,.
  \label{eq::alpha123}
\end{eqnarray}
which holds for any renormalization scale $\mu$. 
We also need the $W$  and $Z$ boson pole masses $M_W$ and $M_Z$,  
the top quark and tau lepton pole
masses $M_t$ and $M_\tau$ and the
running bottom quark mass $\mbMSbar$.
For the convenience of the reader we also specify their numerical
values~\cite{Amsler:2008zzb,:1900yx,Chetyrkin:2009fv}:
\begin{eqnarray}
  M_W &=& 80.398~\mbox{GeV}  \,,\nonumber\\
  M_Z &=& 91.1876~\mbox{GeV}  \,,\nonumber\\
  M_t &=& 173.3~\mbox{GeV}  \,,\nonumber\\
  M_\tau &=& 1.77684~\mbox{GeV}  \,,\nonumber\\
  \mbMSbar(\mbMSbar) &=& 4.163~\mbox{GeV}  \,.
\end{eqnarray}
The corresponding uncertainties are not important for our analysis.
These parameters are converted to six-flavor theory using {\tt RunDec}~\cite{Chetyrkin:2000yt}
and then used to compute the starting values for $y_t, y_b$ and $y_\tau$ at the electroweak scale.
The RGE running in the SM was implemented at two loops~\cite{Jones:1981we,Machacek:1983fi,Machacek:1983tz,Ford:1992pn}
for the electroweak sector and at three loops~\cite{Tarasov:1980au,Larin:1993tp} for QCD.
We take into account the tau, bottom and top Yukawa couplings and thus solve the 
coupled system of six differential equations. Since the quartic SM Higgs 
coupling $b$ enters the equations of the Yukawa couplings
starting from two-loop order only, we neglect its contribution.
After taking into account the two-loop decoupling relations, we compute the running
from $\mugut$ to the Planck scale using three-loop RGEs for the gauge coupling
and one-loop RGEs for the Yukawa couplings.
The RGEs are obtained by inserting the general expressions for
the Yukawa and scalar couplings (eqs. (\ref{eq::su5yuk}) and (\ref{eq::su5lam})) as well as the numerical values
for the group theory factors into the general formulas of refs.~\cite{Pickering:2001aq,Machacek:1983fi}
(see appendix~\ref{sec::betasu5} for the details).

\begin{figure}[t]
  \centering
  \begin{tabular}{c}
    \includegraphics[width=.75\linewidth]{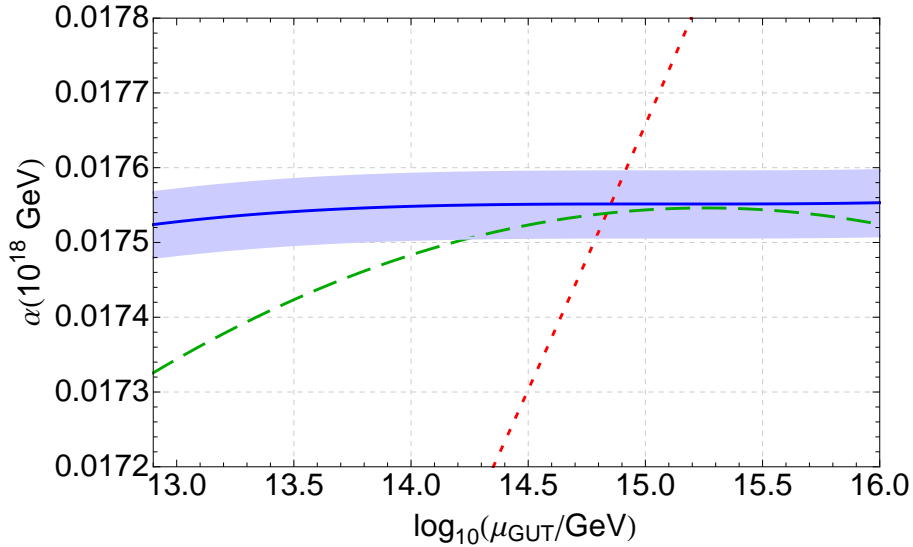}
  \end{tabular}
  \caption{\label{fig::scaledep}
    Dependence of $\alpha(10^{18}\ {\rm GeV})$ on the decoupling scale $\mugut$.
    The red (dotted), green (dashed) and blue (solid) lines correspond
    to the one-, two- and three-loop analysis, respectively.
    For the three-loop curve also the experimental error band with
    $\delta\alpha_s=0.0020$ has been indicated.}
\end{figure}

In figure \ref{fig::scaledep} the dependence on the decoupling scale of $\alpha(10^{18} {\rm GeV})$
is shown.
Since only for QCD the full three-loop $\beta$ function could be implemented
and there is no unification of gauge couplings anyway, we took 
$\alpha(\mugut)=\zeta^{-1}_{\alpha_3}(\mugut)\ \alpha_3(\mugut)$ as a starting value for the gauge coupling above
the GUT scale. 
For illustration we use the following set of mass parameters:
\begin{eqnarray}
 \mx &=& 10^{15}\ {\rm GeV}\,, \nonumber\\
 \mhc &=& 4 \cdot 10^{13}\ {\rm GeV}\,, \nonumber\\
 \msig &=&  10^{14}\ {\rm GeV}\,, \nonumber\\
 M_{24} &=& 6 \cdot 10^{13}\ {\rm GeV}
\end{eqnarray}
which are chosen to obey the restriction 
$\mx \gtrsim M_i$ for $i= {\rm H_c, \Sigma}, 24$ .
Otherwise the scalar self-couplings easily become non-perturbative and blow up
the gauge coupling above the GUT scale. The scale dependence is shown for
$n$-loop running and $(n-1)$-loop decoupling with $n=1,2,3$. 
We observe a dramatic improvement when going from $n=2$
to $n=3$. In particular the three-loop corrections can be larger
than the experimental error band depending on $\mugut$. Note also that for $n=2$
choosing $\mugut$ naively as a mean value of the GUT masses
which would be of $\mathcal O(10^{14}\ {\rm GeV})$ in our case
is not a good choice.
The described qualitative behavior
does not depend much on our choice of the GUT masses. 
Though the numerical effect of the two-loop matching is already significant in the Georgi-Glashow model,
we emphasize that in certain models that contain large representations, as e.g.
the Missing Doublet Model~\cite{Masiero:1982fe,Grinstein:1982um}, 
we expect these corrections to be even larger~\cite{Martens:2010nm}.
Our goal for the future, of course, is to generalize the formula for $\zeta_{\alpha_i}$ to 
make it applicable to these models.

To provide a check of the result for the Georgi-Glashow model,
we have verified analytically that the matching coefficients 
$\zeta_{\alpha_i}(\mugut)$ exhibit the correct $\mugut$ dependence. 
This can be derived from the knowledge of the two-loop $\beta$ functions of the SM
and the $SU(5)$ model by computing the derivative w.r.t. $t_{\rm GUT}\equiv\ln(\mugut)$
of eq.~(\ref{eq::decrelation}). Solving the resulting differential equation
order by order, we arrive at a general formula\footnote{For simplicity we 
neglect the Yukawa corrections in this formula. However, the generalization is straightforward.
Of course, in our analytical check we took care of them too.}
for the $\mugut\,$-dependent terms in $\zeta_{\alpha_i}(\mugut)$:
\begin{eqnarray}
\label{eq::gennudep}
 &&\zeta_{\alpha_i}(\mugut) = 1 + \frac{\alpha(\mugut)}{\pi}\Big[
          \tfrac{1}{2}(\beta_0^i - \beta_0)\, t_{\rm GUT} - C_0(M_{\rm h}) \Big]  \\
       &&+ \left(\frac{\alpha(\mugut)}{\pi}\right)^2 \Bigg[\tfrac{1}{4} (\beta_0^i - \beta_0)^2\, t_{\rm GUT}^2
         + \Big[ \tfrac{1}{8}(\beta_1^i - \beta_1) 
         - C_0(M_{\rm h})\, (\beta_0^i - \beta_0) \Big]\, t_{\rm GUT} + C_1(M_{\rm h}) \Bigg]\,. \nonumber
\end{eqnarray}
$C_0$ and $C_1$ are $\mugut$-independent terms that depend only on the heavy GUT masses.
The $\beta$ function coefficients are defined by:
\begin{eqnarray}
\frac{1}{2} \frac{d}{d t } \frac{\alpha}{4\pi} &=& \sum_{k=0}^{N-1}\, 
    \left(\frac{\alpha}{4\pi}\right)^{k+2}\ \beta_k\,, \qquad 
  \alpha = \frac{g^2}{4\pi}
\end{eqnarray}
and similarly for $\alpha_i$. We find agreement in the $\mugut$ dependence
of our explicit calculation with the 
form of eq.~(\ref{eq::gennudep}).


\section{\label{sec::concl}Conclusions and Outlook}
As experimental accuracy for $\alpha_s$, $\alpha_{em}$ and $\sin\Theta$ is increasing, also theoretical unification analyses
must improve their precision in order to find better exclusion limits for GUTs.
Therefore, we have performed a first step towards the calculation of
the two-loop matching corrections for the gauge couplings at the GUT scale in a framework that aims at making as few
assumptions on the underlying GUT as possible. The assumptions that were made can be found in subsection \ref{subsec::dec}.
The result is general enough to be applied to the simplest GUT, the Georgi-Glashow $SU(5)$.
The numerical impact in this model was found to be larger than the current experimental uncertainty on $\alpha_s$
depending on the choice of the decoupling scale $\mugut$. 
We expect even larger effects in GUT models that contain large representations, 
as the so-called Missing Doublet Model. Moreover, the two-loop matching coefficients provide
a significant stabilization of our predictions w.r.t the variation of the decoupling scale.

Furthermore, we have described in detail the proper 
treatment of the gauge fixing and the renormalization procedure for
this calculation. In this context also the issue of Higgs tadpoles 
in theories with spontaneous symmetry breaking
was discussed in a general manner.  The appendix contains an 
in-depth introduction of the group theoretical framework that we used. 
There we also give many useful reduction identities that are 
essential for performing multi-loop calculations in GUTs and can
also be applied to SUSY GUTs.

We consider the result computed in this paper only as an intermediate 
step towards a more general calculation of $\zeta_{\alpha_i}$. 
Particularly, we are interested in performing a consistent three-loop RGE analysis 
for SUSY GUTs, where two-loop matching at the GUT scale is needed.
In order to apply our result to the simplest SUSY GUT, the minimal SUSY $SU(5)$, several
generalizations are needed:
\begin{itemize}
  \item Add a trilinear term to the scalar potential eq.~(\ref{eq::vphi}).
  \item Add two massive Dirac and three massive Majorana fermion SM-irreducible representations
        as well as all possible kinds of Yukawa interactions for them.
  \item Increase the number of SM-irreducible representations within $\mathcal{R}^\mathcal{H}$ by one
        in order to cover the case of the non-renormalizable version of minimal SUSY $SU(5)$
  \item Allow for unitary mixing matrices in the interactions of heavy Dirac fermions with gauge bosons.
  \item Convert the result to the \drbar~renormalization scheme.
\end{itemize}
Though there is quite some work to be done, to achieve this, 
we have already built a solid basis to start with
and have  overcome many of the main difficulties that were  expected to occur.




\section*{Acknowledgments}
The author is greatly indebted to Luminita Mihaila and Matthias Steinhauser
for countless fruitful discussions and many useful comments to the manuscript.
Thanks also to Jennifer Girrbach for carefully reading the manuscript 
and providing useful remarks.
This work was supported by the ``Studienstiftung des deutschen Volkes'', 
the DFG through SFB/TR~9 ``Computational Particle Physics'' 
and the Graduiertenkolleg ``Hochenergiephysik und Teilchenastrophysik''.
\vspace{2cm}

\begin{appendix}
\begin{center}
 \textbf{\begin{LARGE}Appendix\end{LARGE}}
\end{center}


  \section{\label{sec::group}Group theoretical framework}
When performing multi-loop calculations for GUTs, one necessarily has to deal 
with the reduction of the color structure\footnote{For convenience
we will use the terms ``color structure'', ``color factor'' etc.
in this appendix following QCD terminology. However, obviously our formalism
is not restricted to $SU(3)$, but is meant to be applied to GUT groups.}
~\cite{Binetruy:1980xn,Hall:1980kf,Machacek:1983tz,Machacek:1983fi,Pickering:2001aq,vanDamme:1983fw}.
The aim is to reduce all the color factors of the diagrams to a set of
basic invariants. In this way nontrivial cancellations among different diagrams
are possible without inserting the actual numerical values for those invariants.
In this appendix we develop a notational framework that is appropriate for
this task and also give some useful reduction identities that will hopefully
prove helpful also for future calculations. For similar reduction algorithms and identities
applied to unbroken gauge theories see e.g. refs.~\cite{vanRitbergen:1998pn,Cvitanovic:1976am}
and references therein.
Here we focus on spontaneously broken gauge theories.

We start with a simple GUT group $\mathbf{G}$ that is broken to a (in general not simple)
gauge group $\prod_k \mathbf{G}_k$.  In the following the $\mathbf{G}_k$ are called group factors.
The generators of $\mathbf{G}$ in a general reducible representation $\mathcal R$ are denoted by $T^\alpha$
in this appendix\footnote{In the main text we employed the symbol $T^\alpha$ only for the 
fermion representation. In this appendix, however, we will use the symbol for a generic generator
of the gauge group.}.
They fulfill the commutation relation
\begin{equation}
 [T^\alpha, T^\beta] = i f^{\alpha\beta\gamma}\, T^\gamma\,,
\end{equation}
where $f^{\alpha\beta\gamma}$ are the structure constants of $\mathbf{G}$.
In order to distinguish between broken and unbroken generators, we use the notation
of subsection \ref{subsec::lag}:
\begin{equation}
\label{eq::splitindex2} 
 \{\alpha\} = \sum_i\{A_i\} + \sum_i \{a_i\} = \{A\} + \{a\}
\end{equation}
where $A_i$ label the broken generators of $\mathbf{G}$ belonging to the $\prod_k \mathbf{G}_k$-irreducible subspace labeled by $i$.
If there is only one $\prod_k \mathbf{G}_k$-irreducible subspace in the adjoint representation of $\mathbf{G}$,
we can omit the sub-index $i$ in $A_i$.
In contrast, $a_i$ label the unbroken generators belonging to the subgroup $\mathbf{G}_i$.
As the $\mathbf{G}_i$ are regular subgroups of $\mathbf{G}$, also the following cummutation relations hold:
\begin{equation}\label{eq::commutator}
 [T^{a_i},T^{b_j}] = i f^{a_i b_j c_k}\, T^{c_k}
\end{equation}
where $f^{a_i b_j c_k} =0$ unless $i=j=k$. 
Furthermore, because the subgroup $\mathbf{G}_i$ is closed, we have $f^{a_i b_j A_k}=0$ for all $i,j,k$. 
Otherwise the commutator $[T^{a_i},T^{b_j}]$ would contain terms
proportional to the broken generators $T^{A_k}$.
Note that if not explicitly stated, the sub-indices $i,j...$ 
of the indices $a_i,b_j..., A_i, B_j...$ are not summed.
A repeated index $a,b...$ or $A,B...$ without sub-index 
means that the sub-index has been summed over.
The representation $\mathcal{R}$ that $T^\alpha$ is defined on generally is
reducible under $\mathbf{G}$. It decomposes into $\mathbf{G}$-irreducible representations
where the gauge bosons, fermions and scalars of the theory live in:
\begin{eqnarray}
 \mathcal{R} \rightarrow \bigoplus_x \mathcal{R}^x  = \mathcal{R}^{\mathcal{A}}
     \oplus \mathcal{R}^{\mathcal{H}} 
     \oplus \mathcal{R}^{\mathcal{S}_{\rm I}} 
     \oplus \mathcal{R}^{\mathcal{S}_{\rm II}} \oplus ... \oplus \mathcal{R}^{\mathcal{F}_{\rm I}} 
     \oplus \mathcal{R}^{\mathcal{F}_{\rm II}} \oplus ...\,.
\end{eqnarray}
$\mathcal{R}^{\mathcal{A}}$ stands for the adjoint representation and 
$\mathcal{R}^{\mathcal{H}}$ for the representation of the GUT-breaking scalar.
The other symbols represent the irreducible representations for the 
scalars and fermions, respectively, numbered by roman numerals for convenience.
We define projectors on these subspaces denoted by $\Pi^x$ with 
$x=\mathcal{A},\mathcal{H},\mathcal{S}_{\rm I},\mathcal{S}_{\rm II}...,\mathcal{F}_{\rm I},\mathcal{F}_{\rm II}...$.
Clearly, $[\Pi^x,T^\alpha]=0$ holds for all $x$ and $\alpha$.
Each of those representations decomposes further under $\prod_k \mathbf{G}_k$:
\begin{eqnarray}
 \mathcal{R}^x \rightarrow \bigoplus_n \mathcal{R}^x_n = \mathcal{R}^x_1 \oplus \mathcal{R}^x_2 \oplus \mathcal{R}^x_3 \oplus ...
\end{eqnarray}
with projectors $P^x_n$ and $p^x_n$. We use a capital $P$ to denote that
the respective $\prod_k \mathbf{G}_k$-irreducible representation contains fields
with a mass of $\mathcal O(\mgut)$ and a lowercase $p$ for projectors
on a subspace with massless fields. Because the Lagrangian
is still invariant under $\prod_k \mathbf{G}_k$ after symmetry breaking, 
each of those $\prod_k \mathbf{G}_k$-irreducible subspaces can be assigned
to a definite mass of the respective field.
The indices $n=1,2,3...$ label the $\prod_k \mathbf{G}_k$-irreducible representation
in $\mathcal{R}^x$. The projectors fulfill $[P^x_n,T^{a_i}]=0=[p^x_n,T^{a_i}]$ for all $x,a,n$ and $i$.
Furthermore, we have
\begin{equation}
\sum_n \wP^x_n \equiv \sum_n p^x_n + \sum_n P^x_n \equiv p^x + P^x = \Pi^x \,
\end{equation}
where $\wP$ can 
be $P$ or $p$ depending on $n$.
Armed with these definitions, we can define some basic Dynkin indices $I_2(...)$ and 
Casimir invariants $C_2(...)$ which have real numerical values for a given group.
\begin{eqnarray}
\label{eq::casimirs}
 {\rm Tr}(\Pi^x T^\alpha T^\beta) &=& I_2(\Pi^x)\ \delta^{\alpha\beta} \,,\nonumber\\
 {\rm Tr}(\wP^x_n T^{a_i} T^{b_i}) &=& I_2(\wP^x_n)^i\ \delta^{a_ib_i} \,,\nonumber\\
 {\rm Tr}(\wP^x_n T^{A_i} \wP^x_m T^{B_i}) &=& 
    I_2(\wP^x_n,\wP^x_m)^i\ \delta^{A_i B_i} \,,\nonumber\\
 \Pi^x T^\alpha T^\alpha &=& C_2(\Pi^x)\ \Pi^x \,,\nonumber\\
 \wP^x_n T^{a_i} T^{a_i} &=& C_2(\wP^x_n)^i\ \wP^x_n \,,\nonumber\\
 \wP^x_n T^{A_i} \wP^x_m T^{A_i} &=& 
    C_2(\wP^x_n,\wP^x_m)^i\ \wP^x_n \,.
\end{eqnarray}
Again $\wP$ can stand for either $P$ or $p$.
Furthermore, we can define the dimensions of the various irreducible
representations by
\begin{eqnarray}
\label{eq::dim}
 \Delta^x &\equiv& {\rm Tr}(\Pi^x) \,,\nonumber\\
 D^x_n &\equiv& {\rm Tr}(P^x_n) \,,\nonumber\\
 d^x_n &\equiv& {\rm Tr}(p^x_n)\,,
\end{eqnarray}
and specifically for the adjoint representation:
\begin{eqnarray}
\label{eq::dimadj}
 \Delta^\mathcal{A} &\equiv& \delta^{\alpha\alpha} \,,\nonumber\\
 D^\mathcal{A}_n &\equiv& \delta^{A_nA_n} \,,\nonumber\\
 d^\mathcal{A}_n &\equiv& \delta^{a_na_n} \,.
\end{eqnarray}
which gives us the relations 
\begin{eqnarray}
\label{eq::casrel}
 I_2(\Pi^x)\ \Delta^{\mathcal{A}} &=& C_2(\Pi^x)\ \Delta^x \,,\nonumber\\
 I_2(P^x_n)^i\ d^{\mathcal{A}}_i &=&   C_2(P^x_n)^i\ D^x_n \,,\nonumber\\
 I_2(p^x_n)^i\ d^{\mathcal{A}}_i &=&   C_2(p^x_n)^i\ d^x_n \,,\nonumber\\
 I_2(P^x_n,\wP^x_m)^i\ D^{\mathcal{A}}_i &=& C_2(P^x_n,\wP^x_m)^i\ D^x_n \,,\nonumber\\
 I_2(p^x_n,\wP^x_m)^i\ D^{\mathcal{A}}_i &=& C_2(p^x_n,\wP^x_m)^i\ d^x_n \,.
\end{eqnarray}
Let us emphasize again that no summation over the sub-indices $i$ that label the 
$\prod_k \mathbf{G}_k$-irreducible representation is implied here. 
However, we introduce the 
convention that omitting this sub-index implies summation:
\begin{eqnarray}
 C_2(...) &\equiv& \sum_i C_2(...)^i \,, \nonumber\\
 C_2(\wP^x_n,\wP^x)^i &\equiv&  \sum_m C_2(\wP^x_n,\wP^x_m)^i \,, \nonumber\\
 I_2(...\wP^x ...)^i &\equiv& \sum_m I_2(...\wP^x_m ...)^i \,.
\end{eqnarray}
Note also that the invariants $I_2(\Pi^x)$ and $C_2(\Pi^x)$ have only been 
defined for convenience. Actually they can be decomposed into other
``more elementary'' invariants:
\begin{eqnarray}
I_2(\Pi^x) &=&  I_2(p^x)^i + I_2(P^x)^i 
   = I_2(p^x,p^x)^i + I_2(P^x,P^x)^i  + 2\, I_2(P^x,p^x)^i    \,,  \\
C_2(\Pi^x) &=&  C_2(p^x_i) + C_2(p^x_i,p^x) + C_2(p^x_i,P^x)
 =   C_2(P^x_i) + C_2(P^x_i,p^x) + C_2(P^x_i,P^x) \,,\nonumber
\end{eqnarray}
where the summation convention introduced before has been used. The right hand side
of these equations does not depend on $i$ anymore due to the summation of the projectors
over the full representation space $\mathcal{R}^x$.
Since in the calculation of matching coefficients one has to distinguish between heavy
and light particles, it is more convenient to use the ``more elementary'' invariants
on the right hand side.

The definitions that have been introduced are general enough to be applied to all 
color factors that appear in our calculations. However, some of the fields
live in representations that deserve special attention.
First let us focus on the adjoint representation. The generators here
are defined by
\begin{equation}
 (T^\alpha_{\mathcal{A}})_{\beta\gamma} \equiv 
  (\Pi^{\mathcal{A}}\, T^\alpha)_{\beta\gamma} = -i f^{\alpha\beta\gamma} \,.
\end{equation}
Clearly, the operators $P^{\mathcal{A}}_i$ and $p^{\mathcal{A}}_i$ project 
on the subspaces with indices $A_i$ and $a_i$, respectively.
Because now $x=\mathcal{A}$ in eq.~(\ref{eq::casrel}) and $f^{a_ib_jA_k}=0$, things
simplify for the adjoint representation. In fact it is sufficient to define
four invariants:
\begin{eqnarray}
\label{eq::adjinv}
 f^{\alpha\gamma\delta}f^{\beta\gamma\delta} &=& I_2(\Pi^{\mathcal{A}})\ \delta^{\alpha\beta} \,,\nonumber\\
 f^{a_i c_i d_i}f^{b_i c_i d_i} &=& I_2(p^{\mathcal{A}}_i)^i \ \delta^{a_i b_i} \,,\nonumber\\
 f^{a_i C_j D_j}f^{b_i C_j D_j} &=& I_2(P^{\mathcal{A}}_j)^i\ \delta^{a_ib_i} \,,\nonumber\\
 f^{A_j c_i D_j}f^{B_j c_i D_j} &=& I_2(P^{\mathcal{A}}_j,p^{\mathcal{A}}_i )^j \ \delta^{A_jB_j} 
\end{eqnarray}
where we followed the notation of eq.~(\ref{eq::casimirs}). We keep the redundant sub-indices
$i$ and $j$ in lines 2 and 4 of this equation in order to be consistent with the notation of 
eq.~(\ref{eq::casimirs}). 
Note that $f^{a_i B_j C_k} = 0$ for $j \neq k$ because the indices $j$ and $k$ are used
here to label the $\prod_k \mathbf{G}_k$-irreducible representation of the generator $T^{a_i}_{\mathcal{A}}$.
There is another quadratic Casimir invariant for the adjoint representation
that can be expressed through these and one that vanishes:
\begin{eqnarray}
   f^{A_j C D}f^{B_j C D} &=& \Big[I_2(\Pi^{\mathcal{A}})\ 
    - 2\,  I_2(P^{\mathcal{A}}_j,p^{\mathcal{A}} )^j\Big]\ \delta^{A_jB_j} \,, \nonumber\\
   f^{A_i C D}f^{b_j C D} &=& 0 \,.
\end{eqnarray}
Furthermore, there are relations between these invariants:
\begin{eqnarray}
  I_2(\Pi^{\mathcal{A}}) &=& I_2(P^{\mathcal{A}})^i + I_2(p^{\mathcal{A}}_i)^i \,, \nonumber\\
 I_2(P^{\mathcal{A}}_j,p^{\mathcal{A}}_i )^j\, D^{\mathcal{A}}_j &=& I_2(P^{\mathcal{A}}_j)^i\, d^{\mathcal{A}}_i \,.
\end{eqnarray}
At the two-loop level these relations are not sufficient to reduce all adjoint color factors
because products of up to six structure constants with various contractions can appear in the diagrams.
Using the Jacobi identity for $f^{\alpha\beta\gamma}$, we can derive relations for
products of three contracted structure constants
that have three open indices:
\begin{eqnarray}
\label{eq::fffrelations}
 f^{\alpha\delta\epsilon} f^{\beta\epsilon\phi}f^{\gamma\phi\delta} &=&  
     \frac{1}{2} I_2(\Pi^{\mathcal{A}})\, f^{\alpha\beta\gamma} \,,\nonumber\\
 f^{a_i d_i e_i}f^{b_i e_i f_i}f^{c_i f_i d_i} &=&  
     \frac{1}{2} I_2(p^{\mathcal{A}}_i)^i \, f^{a_i b_i c_i} \,,\nonumber\\
  f^{a_i D E}f^{b_i E F}f^{c_i F D} &=&  
     \frac{1}{2} I_2(P^{\mathcal{A}})^i \, f^{a_i b_i c_i} \,,\nonumber\\
  f^{a_i D E}f^{b_i E F}f^{C F D} &=& 0 \,,\nonumber\\
  f^{a_i D E}f^{B_j E F}f^{C_j F D} &=&  
    \frac{1}{2}\Big[I_2(\Pi^{\mathcal{A}}) - 2\, 
     I_2(P^{\mathcal{A}}_j,p^{\mathcal{A}} )^j \Big]\, f^{a_i B_j C_j} \,,\nonumber\\
  f^{a_i d_i e_i}f^{B_j e_i F}f^{C_j F d_i} &=&  \frac{1}{2} I_2(p^{\mathcal{A}}_i)^i \,  f^{a_i B_j C_j} \,,\nonumber\\
  f^{a_i D E}f^{B_k E f_j}f^{C_k f_j D} &=&  
     \frac{1}{2}\Big[ 2\, I_2(P^{\mathcal{A}}_k,p^{\mathcal{A}}_j )^k  
    - \delta_{ij}\, I_2(p^{\mathcal{A}}_i)^i \Big] f^{a_i B_k C_k} \,,\nonumber\\
  f^{A_j D e}f^{B_j e F}f^{C_j F D} &=&  
     \frac{1}{2}\,  I_2(P^{\mathcal{A}}_j,p^{\mathcal{A}} )^j f^{A_j B_j C_j} \,,\nonumber\\
  f^{A_j D E}f^{B_j E F}f^{C_j F D} &=&   
     \frac{1}{2}\Big[I_2(\Pi^{\mathcal{A}}) - 3\, 
      I_2(P^{\mathcal{A}}_j,p^{\mathcal{A}} )^j \Big] f^{A_j B_j C_j} \,.
\end{eqnarray}
These relations are sufficient to do all the reduction for two-point and three-point Green's functions
for the adjoint representation at the two-loop level.

Next, let's turn our attention to the representation $\mathcal{R}^\mathcal{H}$
where the GUT-breaking scalar field $H$ and the Goldstone field $G$ live in.
Some peculiarities occur here due to the appearance of the vev $v$. 
$\mathcal{R}^\mathcal{H}$ decomposes under $\prod_k \mathbf{G}_k$ into the part of physical 
Higgs fields and a part of Goldstone bosons:
\begin{eqnarray}
 \mathcal{R}^\mathcal{H} \rightarrow \mathcal{R}^\mathcal{\tilde {H}} 
  \oplus \mathcal{R}^\mathcal{G} = \mathcal{R}^\mathcal{\tilde {H}}_1 \oplus
 \mathcal{R}^\mathcal{\tilde {H}}_2 \oplus...  \oplus   \mathcal{R}^\mathcal{G}_1 
 \oplus \mathcal{R}^\mathcal{G}_2 \oplus... 
\end{eqnarray}
As already explained in subsection~\ref{subsec::lag},
the explicit form of the projector on the subspace $\mathcal{R}^\mathcal{G}$
is given by~\cite{Langacker:1980js}:
\begin{eqnarray}
 \label{eq::Gproj}
 P^\mathcal{G}_{i} = g^2\tilde{T}^{A_i} v \left(\frac{1}{M^2_X}\right)_{A_iB_i} v\tilde{T}^{B_i} \,, 
 \qquad  P^\mathcal{G} = \sum_i  P^\mathcal{G}_{i}
\end{eqnarray}
where the (diagonal) gauge boson mass matrix
\begin{equation}
 \label{eq::Xmassmatrix}
 (\mx^2)_{A_iB_i} \equiv g^2 v \tilde{T}^{A_i} \tilde{T}^{B_i}  v \equiv M_{{\rm X}_i}^2 \delta^{A_iB_i}
\end{equation}
has been used. 
The antisymmetric generators in the real representation $\mathcal{R}^{\mathcal{H}}$ 
have been denoted by $\tilde{T}^\alpha$. 
Accordingly, the projectors on the subspace of physical
Higgs bosons can be written as
\begin{eqnarray}
 \sum_i  P^\mathcal{\tilde{H}}_i = P^\mathcal{\tilde{H}} \equiv \Pi^\mathcal{H} - P^\mathcal{G} =  
  \Pi^\mathcal{H} - g^2\tilde{T}^{A} v \left(\frac{1}{M^2_X}\right)_{AB} v\tilde{T}^{B} \,.
\end{eqnarray}
where $P^\mathcal{\tilde{H}}_i$ projects on $\mathcal{R}^\mathcal{\tilde{H}}_i$.
With these definitions and using eq.~(\ref{eq::commutator}) 
as well as the antisymmetry of $\Tt^\alpha$, we already can derive a useful reduction identity for 
an invariant tensor that appears frequently:
\begin{equation}
\label{eq::mxfabc}
 v \tilde{T}^{A_i} \tilde{T}^{B_j} \tilde{T}^{C_k} v = 
  \frac{i}{2g^2} ( M_{{\rm X}_i}^2 - M_{{\rm X}_j}^2 + M_{{\rm X}_k}^2 ) f^{A_iB_jC_k} \,.
\end{equation} 
One important property follows from $\Tt^{a_i} v = 0$:
\begin{equation}
 \Tt^{a_i} \Tt^{A_j} v = [\Tt^{a_i}, \Tt^{A_j}]\, v = - (T^{a_i}_{\mathcal{A}})_{A_jB_j} \Tt^{B_j} v \,.
\end{equation}
i.e. $\Tt^{a_i}$ acts like the adjoint generator on the subspace of Goldstone bosons.
This leads to various relations between invariants in $\mathcal{R}^{\mathcal{H}}$
and in $\mathcal{R}^{\mathcal{A}}$:
\begin{eqnarray}
 D^\mathcal{G}_j &=& D^\mathcal{A}_j  \,, \nonumber\\
 I_2(P^\mathcal{G}_j)^i &=&  I_2(P^\mathcal{A}_j)^i  \,, \nonumber\\
 C_2(P^\mathcal{G}_j)^i &=&  I_2(P^\mathcal{A}_j,p^\mathcal{A}_i)^j  \,, \nonumber\\
 I_2(P^{\mathcal{\tilde{H}}})^i &=& I_2(\Pi^\mathcal{H}) -  I_2(P^\mathcal{A})^i \,,\nonumber\\
 I_2( P^{\mathcal{\tilde{H}}},P^{\mathcal{\tilde{H}}})^i &=& I_2(\Pi^{\mathcal{H}}) 
     - 2\ C_2(\Pi^{\mathcal{H}}) + \frac{3}{2}\ I_2(P^\mathcal{A}_i,p^\mathcal{A})^i  
     + \frac{1}{4}\ I_2(\Pi^\mathcal{A}) \,. 
\end{eqnarray}
There are also two nontrivial important reduction identities that involve
both types of invariants:
\begin{eqnarray}
\label{eq::redfaBCtr1}
 && f^{a_iB_jC_j}\ {\rm Tr}(\wP^{x}_n T^{B_j}
    \wP^{x}_m T^{C_j} T^{b_i})  = \nonumber\\
  &&  \frac{i}{2}\delta^{a_ib_i} \Big[ I_2(P^{\mathcal{A}}_j)^i I_2(\wP^{x}_n,\wP^{x}_m )^j 
  + C_2( \wP^{x}_n,\wP^{x}_m )^j I_2(\wP^{x}_n)^i 
  - C_2(\wP^{x}_m,\wP^{x}_n)^j I_2(\wP^{x}_m)^i\Big] \,,\\[0.5cm]
\label{eq::redfaBCtr2}
 && \sum_{jk}\ f^{a_iB_jC_j}f^{b_iD_kE_k}\,
      v\tilde{T}^{B_j}\tilde{T}^{D_k} P^\mathcal{\tilde{H}}_n \tilde{T}^{C_j}\tilde{T}^{E_k}v =\nonumber\\
&& \delta^{a_ib_i}\sum_j \frac{M_{{\rm X}j}^2}{g^2} 
    \Big[ I_2(P^\mathcal{A}_j)^i\,  I_2(P^{\mathcal{\tilde{H}}}_n,P^\mathcal{G})^j 
  - \frac{1}{2} I_2(P^\mathcal{\tilde{H}}_n)^i\, C_2(P^{\mathcal{\tilde{H}}}_n,P^\mathcal{G})^j \Big] \,
\end{eqnarray}
where $x\in\{\mathcal{\tilde{H}},\mathcal{S_{\rm I}},\mathcal{S_{\rm II}},
...,\mathcal{F_{\rm I}},\mathcal{F_{\rm II}}...\}$.
Additionally, w.l.o.g. we now define $P^\mathcal{\tilde{H}}_1$ to be the operator that
projects on the subspace where $v\neq 0$ (i.e. $(P^\mathcal{\tilde{H}}_1)_{ij}=\frac{v_iv_j}{v^2}$
is a matrix with a single non-zero entry in the component $(k,k)$ where $v_k\neq 0$).
Then because of $\Tt^{a_i} v = 0$
and $v\Tt^{A_i} P^\mathcal{\tilde{H}}_n = v\Tt^{A_i} P^\mathcal{G} P^\mathcal{\tilde{H}}_n = 0$, 
any invariant that contains $P^\mathcal{\tilde{H}}_1$ alone or together with
some other $P^\mathcal{\tilde{H}}_i$ vanishes:
\begin{eqnarray}
 C_2(P^\mathcal{\tilde{H}}_1) = I_2(P^\mathcal{\tilde{H}}_1)=
 C_2(P^\mathcal{\tilde{H}}_1,P^\mathcal{\tilde{H}}_i) = I_2(P^\mathcal{\tilde{H}}_1,P^\mathcal{\tilde{H}}_i) = 0 \,.
\end{eqnarray}
Since the Higgs mass matrix $M^2_H$ (cf. eq.~(\ref{eq::higgsmass}))
commutes with all $\Tt^{a_i}$, it is
diagonal and proportional to the unit matrix on each $\prod_k \mathbf{G}_k$-irreducible
subspace. Therefore, it can be written as:
\begin{eqnarray}
 M_H^2 = \sum_i P^\mathcal{\tilde{H}}_i\, M^2_{H_i} \,,
\end{eqnarray}
where $M^2_{H_i}$ are masses of the physical Higgs bosons
and particularly $P^\mathcal{G}\, M_H^2 = 0$ due to the Goldstone theorem.

Next we will give some useful reduction identities that involve the 
quartic scalar coupling 
$\lambda^\mathcal{H}_{ijkl} \equiv \lambda_{i'j'k'l'}\,P^\mathcal{H}_{i'i}\,
P^\mathcal{H}_{j'j}\,P^\mathcal{H}_{k'k}\,P^\mathcal{H}_{l'l} $, which is a
totally symmetric invariant tensor under $\mathbf{G}$. These identities can be derived by 
using eq.~(\ref{eq:lamgaugeinv}) and Schur's Lemma as well as the 
definition of the Higgs mass matrix in eq.~(\ref{eq::higgsmass}).
Some of these identities can be obtained by multiplying eq.~(\ref{eq:lamgaugeinv})
by $\Phi_i\Phi_j\Phi_k\Phi_l$ and performing derivatives w.r.t $\Phi_m$.
They are used to eliminate the coupling $\lambda^\mathcal{H}_{ijkl}$
from the result by expressing it through the physical Higgs masses.
\begin{eqnarray}
 \lambda^\mathcal{H}_{ijkl}\ v_k v_m \tilde{T}^A_{ml} 
   &=& (M_H^2\tilde{T}^A)_{ij} -  (\tilde{T}^A M_H^2)_{ij} \,, \nonumber\\
\Rightarrow \lambda^\mathcal{H}_{ijkl}\ v_j v_m \tilde{T}^A_{mk} v_n \tilde{T}^B_{nl}  
   &=&  -  (v \tilde{T}^B\tilde{T}^A M^2_H)_{i} \,,\nonumber\\
\Rightarrow \lambda^\mathcal{H}_{ijkl}\ v_i v_m 
      \tilde{T}^A_{mj} v_n \tilde{T}^B_{nk} v_r \tilde{T}^C_{rl} &=& 0 \,,\nonumber\\[0.5cm]
\lambda^\mathcal{H}_{ijkl} v_m \tilde{T}^A_{mi} v_n \tilde{T}^B_{nj} v_r \tilde{T}^C_{rk} v_s \tilde{T}^D_{sl} 
  &=&  v \tilde{T}^D\tilde{T}^C M^2_H \tilde{T}^A\tilde{T}^B v  \,,\nonumber\\
&& +\ v \tilde{T}^D\tilde{T}^B M^2_H \tilde{T}^A\tilde{T}^C v  \,,\nonumber\\
&& +\ v \tilde{T}^C\tilde{T}^B M^2_H \tilde{T}^A\tilde{T}^D v  \,,\nonumber\\
\lambda^\mathcal{H}_{ijkl}\lambda^\mathcal{H}_{mjkl}v_iv_m 
   &=& \frac{v^2}{\Delta^\mathcal{H}} \lambda^\mathcal{H}_{ijkl}\lambda^\mathcal{H}_{ijkl}  \,,\nonumber\\
\lambda^\mathcal{H}_{klmn} (v\tilde{T}^{A_i})_k (v \tilde{T}^{B_j})_l (P^\mathcal{\tilde{H}}_s)_{mn} 
   &=& - \lambda^\mathcal{H}_{klmn} (v \Tt^{A_i} \Tt^{B_j})_k  v_l (P^\mathcal{\tilde{H}}_s)_{mn}  \,,\nonumber\\
&& + 2\, \text{Tr}( P^\mathcal{\tilde{H}}_s M_H^2\tilde{T}^{B_j}\tilde{T}^{A_i}) 
  - 2\, \text{Tr}( M_H^2\tilde{T}^{B_j}P^\mathcal{\tilde{H}}_s \tilde{T}^{A_i})  \,,\nonumber\\
\lambda^\mathcal{H}_{ijkl}v_jv_kv_l &=& \frac{3 v_i}{ v^2} v M_H^2 v  \,,\nonumber\\
 \lambda^\mathcal{H}_{ijkk} &=& \delta_{ij}\Big[ \frac{2\ \text{Tr}(M^2_H)}{v^2} 
   + \frac{v M_H^2  v}{ v^4} D^\mathcal{H}  \Big]  \,,\nonumber\\
 \lambda^\mathcal{H}_{ijkl}v_iv_j (P^\mathcal{H}_s)_{kl} 
   &=&  2\, \text{Tr}(M_{H_s}^2 P^\mathcal{H}_s) 
   + \frac{v M_H^2 v}{v^2} D^\mathcal{H}_s \,.
\end{eqnarray}

In the same way we obtain some important relations for the coupling $\lambda$ that is not 
restricted to the space $\mathcal{R}^\mathcal{H}$:
\begin{eqnarray}
\lambda_{ijk'l'}(v\tilde{T}^A)_{i} v_j (\wP^\mathcal{S})_{k'k}\ 
 (\wP^\mathcal{S})_{l'l}&=& (M_S^2\tilde{T}^A)_{kl} - (\tilde{T}^A M_S^2)_{kl} \,,\nonumber\\
\lambda_{ijkl} (v\tilde{T}^A)_{i} (v\tilde{T}^A)_{j}\ (\tilde{T}^{a_n}\tilde{T}^{a_n}\wP^\mathcal{S}_m)_{kl} &=& 
     2\, \text{Tr}(M_S^2 \tilde{T}^A \tilde{T}^A\tilde{T}^{a_n}\tilde{T}^{a_n}\wP^\mathcal{S}_m)  \,,\nonumber\\
     &&-\ 2\, \text{Tr}( \tilde{T}^A M_S^2 \tilde{T}^A\tilde{T}^{a_n}\tilde{T}^{a_n}\wP^\mathcal{S}_m) \nonumber\\
     &&- C_2(P^\mathcal{H})\, \lambda_{ijkl}v_iv_j
      (\tilde{T}^{a_n}\tilde{T}^{a_n}\wP^\mathcal{S}_m)_{kl} \,,\nonumber
\end{eqnarray}
\begin{eqnarray}
\lambda_{i_1i_2i_3i_4} \lambda_{j_1j_2j_3j_4} & v_{i_1} v_{j_1}&
     (\wP^x_{n_1})_{i_2j_2} (\wP^y_{n_2}T^{a_k})_{i_3j_3} (\wP^z_{n_3}T^{a_k})_{i_4j_4} = \nonumber\\
&&\tfrac{1}{2} \lambda_{i_1i_2i_3i_4} \lambda_{j_1j_2j_3j_4} v_{i_1} v_{j_1} (\wP^x_{n_1})_{i_2j_2} 
       (\wP^y_{n_2})_{i_3j_3} (\wP^z_{n_3}T^{a_k}T^{a_k})_{i_4j_4} \nonumber\\
&&- \tfrac{1}{2} \lambda_{i_1i_2i_3i_4} \lambda_{j_1j_2j_3j_4} v_{i_1} v_{j_1} (\wP^y_{n_2})_{i_2j_2} 
       (\wP^z_{n_3})_{i_3j_3} (\wP^x_{n_1}T^{a_k}T^{a_k})_{i_4j_4} \nonumber\\
&&- \tfrac{1}{2} \lambda_{i_1i_2i_3i_4} \lambda_{j_1j_2j_3j_4} v_{i_1} v_{j_1} (\wP^x_{n_1})_{i_2j_2} 
       (\wP^z_{n_3})_{i_3j_3} (\wP^y_{n_2}T^{a_k}T^{a_k})_{i_4j_4} 
\end{eqnarray}

with $x,y,z \in \{\mathcal{H}, \mathcal{S_{\rm I}},\mathcal{S_{\rm II}},... \}$.
In the same way we can also derive reduction identities for the Yukawa coupling $Y^n$
using eq.~(\ref{eq::yukinv})
{\allowdisplaybreaks  
\begin{eqnarray}
\label{eq::yukred}
 {\rm Tr}(\wP^x_j Y^n \wP^y_k Y^{m\star}) (\wP^z_l T^{a_i}T^{a_i})_{nm} &=& 
 {\rm Tr}(\wP^x_j Y^{n\star} \wP^y_k Y^{m} T^{a_i}T^{a_i}) (\wP^z_l)_{nm} \\
 &&+ {\rm Tr}(\wP^y_k Y^{n\star} \wP^x_j Y^{m} T^{a_i}T^{a_i}) (\wP^z_l)_{nm} \nonumber\\
 &&+ {\rm Tr}(\wP^x_j Y^{n\star} \wP^y_k T^{a_i\star} Y^{m}T^{a_i}) (\wP^z_l)_{nm} \nonumber\\
 &&+ {\rm Tr}(\wP^y_k Y^{n\star} \wP^x_j T^{a_i\star} Y^{m}T^{a_i}) (\wP^z_l)_{nm} \,,\nonumber\\
 {\rm Tr}(\wP^x_jT^{a_i} Y^{n\star} \wP^y_k Y^{m}) (\wP^z_l T^{a_i})_{nm} &=& 
 {\rm Tr}(\wP^x_j Y^{n\star} \wP^y_k Y^{m} T^{a_i}T^{a_i}) (\wP^z_l)_{nm} \nonumber\\
 &&+ {\rm Tr}(\wP^y_k Y^{n} \wP^x_j T^{a_i} Y^{m\star}T^{a_i\star}) (\wP^z_l)_{nm} \,,\nonumber\\
 {\rm Tr}(\wP^x_jT^{a_i} Y^{n} \wP^y_k Y^{m\star}) (\wP^z_l T^{a_i})_{nm} &=& 
 -{\rm Tr}(\wP^x_j Y^{n} \wP^y_k Y^{m\star} T^{a_i}T^{a_i\star}) (\wP^z_l)_{nm} \nonumber\\
 &&- {\rm Tr}(\wP^y_k Y^{n\star} \wP^x_j T^{a_i} Y^{m}T^{a_i}) (\wP^z_l)_{nm} \,,\nonumber\\
 {\rm Tr}(\wP^x_j T^{a_i} T^{a_i} Y^{n\star} \wP^y_k Y^{m}) (v\Tt^{A_l})_n (v\Tt^{B_l})_m  &=& 
 - {\rm Tr}(\wP^x_j T^{a_i} T^{a_i} T^{A_l} Y^{n\star} \wP^y_k Y^{m}T^{B_l}) v_n v_m  \nonumber\\
 &&- {\rm Tr}(\wP^x_j T^{a_i} T^{a_i} Y^{n}T^{B_l} \wP^y_k T^{A_l} Y^{m\star}) v_n v_m  \nonumber\\
 &&- {\rm Tr}(\wP^x_j T^{a_i} T^{a_i} T^{A_l} Y^{n\star} \wP^y_kT^{B_l\star} Y^{m}) v_n v_m  \nonumber\\
 &&- {\rm Tr}(\wP^x_j T^{a_i} T^{a_i} T^{B_l\star} Y^{n} \wP^y_k T^{A_l} Y^{m\star}) v_n v_m \,. \nonumber
\end{eqnarray}
}
Here $x,y \in \{\mathcal{F_{\rm I}},\mathcal{F_{\rm II}},... \}$ and 
$z \in \{\mathcal{H}, \mathcal{S_{\rm I}},\mathcal{S_{\rm II}},... \}$. 
The list of reduction identities may not be exhaustive, but it contains
the most important relations. In a similar fashion also other identities can be derived
using the invariance relations eqs. (\ref{eq:lamgaugeinv}) and (\ref{eq::yukinv})
for the invariant tensors.

In the following we briefly sketch the algorithm that 
is used for the reduction of all the color factors in the 
diagrams: we have written a {\tt FORM} program that
treats the color factors for each individual diagram
and reduces them to a basic set of invariants.
In a first step all reduction identities that involve
the quartic scalar coupling $\lambda_{ijkl}$ are applied
to a given expression. After that any expression will contain traces of
strings of generators $T^{a_i}$, $T^{A_i}$, projectors $\wP^x_n$
and Yukawa matrices $Y^n$. The adjoint indices $a_i$ and $A_i$ are either 
contracted with each other or with some structure constants.
First all the contracted adjoint indices are removed, i.e. traces of the form
\begin{equation}
 {\rm Tr}(...T^{a_i}...T^{a_i}...), \qquad {\rm Tr}(...T^{A_i}...T^{A_i}...),
\end{equation}
by applying the definitions of the quadratic Casimir invariants eq.~(\ref{eq::casimirs}).
If the generators are not next to each other, we commute them
until they are and eventually arrive 
at traces that contain no more contracted indices.
Next, expressions of the form 
\begin{equation}
 f^{c_i a_i b_i}\, {\rm Tr}(...T^{a_i}...T^{b_i}...), \qquad f^{\alpha A_i B_i}\, {\rm Tr}(...T^{A_i}...T^{B_i}...),
\end{equation}
are reduced by using
\begin{equation}
 f^{c_i a_i b_i} T^{a_i} T^{b_i}\, = \tfrac{1}{2} f^{c_i a_i b_i}\, [T^{a_i} T^{b_i}]
  = \tfrac{i}{2} I_2(p^\mathcal{A}_i)^i\, T^{c_i}
\end{equation}
and eq.~(\ref{eq::redfaBCtr1}). Again generators that are not next to 
each other are commuted.
Expression that contain the projector $P^\mathcal{G}_i$ are treated separately.
We insert the explicit form of $P^\mathcal{G}_i$ (eq.~(\ref{eq::Gproj}))
and write the traces in the form $v...v$, where ``$...$'' stands for a string
of generators $T^{a_i}$, $T^{A_i}$, and projectors $\wP^x_n$.
Here we additionally make use of the relations 
$T^{a_i} v = 0$ and $ P^\mathcal{\tilde{H}}_i v = 0$ (for $i\neq 1$) in order
to eliminate all generators inside the string that have a lowercase adjoint index.
After that all color factors involving the Yukawa matrix $Y^n$ are reduced
to a basic set of invariants using eq.~(\ref{eq::yukred}). Finally,
we are left with various contractions of structure constants which are
expressed by the respective invariants using eqs. (\ref{eq::adjinv}) and
(\ref{eq::fffrelations}).
The actual program is slightly more complicated than described above
and one needs to introduce repetitive control structures
because not all the reduction can be done by a single run.
However, the basic procedure is as described.

\section{Reparametrization of the scalar potential}\label{sec::reparvphi}
In this appendix give some details on how the parametrization
of the up to quadratic terms in the scalar potential (eq.~(\ref{eq::vphiexp}))
arises. We start with eq.~(\ref{eq::vphi}) and insert the decomposition
of the scalar field $\Phi$:
\begin{eqnarray}
 \Phi_i &=& \underbrace{v_i + H_i} + \underbrace{G_i} + \underbrace{S_i} \\
   &&\underbrace{\hspace{0.4cm}\mathcal{R}^\mathcal{\tilde{H}}
      \hspace{0.9cm}\mathcal{R}^\mathcal{G}}
      \hspace{0.6cm}\mathcal{R}^\mathcal{S} \nonumber\\
   && \hspace{1.0cm}\mathcal{R}^\mathcal{H} \nonumber
\end{eqnarray}
where $v$ is the vev with a single non-zero component, $H$ the physical Higgs field, $G$ the Goldstone field
and $S$ a field, representing all the other scalars, present in the theory.
For clarity we have also indicated the representations where the individual fields live in.
Considering only the up to quadratic terms, we arrive at the following expression:
\begin{eqnarray}\label{eq::vphiinterm1}
 V(\Phi) = H_i \tilde{t}_i + \frac{1}{2}\tilde{M}^2_{ij}H_iH_j + \frac{1}{2}\tilde{M}^2_{ij} G_iG_j
                   + \frac{1}{2}\tilde{M}^2_{ij} S_iS_j + \mathcal{O}(\Phi^3)
\end{eqnarray}
where we have defined the quantities
\begin{eqnarray}
 \tilde{t}_i &=& - \mu^2_{ij}v_j + \frac{1}{6}\lambda_{ijkl}v_jv_kv_l \,, \nonumber\\
 \tilde{M}^2_{ij} &=& - \mu^2_{ij} + \frac{1}{2}\lambda_{ijkl}v_kv_l \,.
\end{eqnarray}
Here we already have used the fact that, due to gauge invariance (eq.~(\ref{eq:lamgaugeinv})),
the matrix $\mu^2$ is proportional to the unit matrix
on each $\mathbf{G}$-irreducible subspace and therefore
\begin{eqnarray}
\mu^2_{ij}v_iG_j &=& 0\,, \qquad \mu^2_{ij}v_iS_j = 0\,,\qquad \mu^2_{ij}H_iG_j = 0 \,, \nonumber\\
\mu^2_{ij}H_iS_j &=& 0\,, \qquad \mu^2_{ij}G_iS_j = 0\,. 
\end{eqnarray}
Also due to gauge invariance (eq.~(\ref{eq:lamgaugeinv})) we have 
$\lambda_{ijkl}v_jv_kv_l\sim v_i$, because the matrix $K$ defined by
 $K_{ij} = \lambda_{ijkl}v_kv_l$ is diagonal and proportional
to the unit matrix on each SM-irreducible subspace. Hence
\begin{eqnarray}
 \lambda_{ijkl}G_iv_jv_kv_l = &0& = 
 \lambda_{ijkl}S_iv_jv_kv_l 
\end{eqnarray}
since $(P^\mathcal{G})_{ij}v_j = 0 = (P^\mathcal{S})_{ij}v_j$.
The parametrization in eq.~(\ref{eq::vphiinterm1}) has the disadvantage
that the masslesness of Goldstone bosons is not manifest there
due to the appearance of an explicit mass matrix for the Goldstone field.
Note furthermore that on the subspace $\mathcal{R}^\mathcal{H}$ the parameter
$\mu^2$ is redundant, because it depends on the tadpole term $\tilde{t}$, which
has been chosen as a physical parameter in the Lagrangian.
Therefore we want $\mu^2$ also to disappear from the mass matrix of the physical
Higgs bosons $H$ by trading it for $\tilde{t}$. To solve these issues, we first rewrite the tadpole term by
observing that $\tilde{t}\sim v_i$ due to arguments already given above. Therefore
we define
\begin{eqnarray}
 \tilde{t}_i \equiv t\, v_i\,,\hspace{1.5cm} t &\equiv& \frac{1}{v^2}\left(-\mu^2_{ij}v_iv_j 
                  + \tfrac{1}{6}\lambda_{ijkl}v_iv_jv_kv_l \right) \nonumber\\
             &\equiv& -\mu_\mathcal{H}^2 + \frac{1}{6v^2}\ \lambda^\mathcal{H}_{ijkl}v_i v_j v_k v_l \,.
\end{eqnarray}
The classical minimum of the potential can be found by setting $t=0$. In a quantum theory, however, 
we must keep $t$ as a counterterm, as explained in subsections~\ref{subsec::lag} and \ref{subsec::ren}.
Next we turn to the term $\frac{1}{2}\tilde{M}^2_{ij} G_iG_j$ in eq.~(\ref{eq::vphiinterm1}). 
From the Goldstone theorem it follows that this mass term must vanish at the classical level. But since we
are also interested in quantum corrections, care is needed here.
We calculate the term by applying the explicit form of the Goldstone projector $P^\mathcal{G}$
which is given in eq.~(\ref{eq::PG}) to $\tilde{M}^2$. Using also the gauge invariance relations
in eq.~(\ref{eq:lamgaugeinv}), we can show the identities
\begin{eqnarray}
 (P^\mathcal{G})_{ij}\lambda_{jklm} v_lv_m
    &=& \frac{1}{3v^2}(P^\mathcal{G})_{ik}\lambda_{jlmn}v_jv_lv_mv_n \,, \nonumber\\
 (P^\mathcal{G})_{ij}\mu^2_{jk} 
   &=& \frac{1}{v^2} (P^\mathcal{G})_{ik}\mu^2_{jl}v_jv_l
    = (P^\mathcal{G})_{ik}\, \mu^2_\mathcal{H} \,.
\end{eqnarray}
Therefore the following identity holds:
\begin{eqnarray}\label{eq::goldstonetheorem}
 (P^\mathcal{G})_{ij} \tilde{M}^2_{jk} = t\, (P^\mathcal{G})_{ik} \,.
\end{eqnarray}
As expected from the Goldstone theorem, the term vanishes at the classical level,
but contributes as a counterterm to the two-point function of the Goldstone field via $t$.
Similarly, we calculate $\frac{1}{2}\tilde{M}^2_{ij} H_iH_j$ by applying
the projector $P^\mathcal{\tilde{H}}=\Pi^\mathcal{H}-P^\mathcal{G}$
to $\tilde{M}^2$:
\begin{eqnarray}
 (P^\mathcal{\tilde{H}})_{ij}\tilde{M}^2_{jk} &=& 
 (\Pi^\mathcal{H})_{ij}\tilde{M}^2_{jk} - (P^\mathcal{G})_{ij}\tilde{M}^2_{jk}\nonumber\\
 &=& t\, (P^\mathcal{\tilde{H}})_{ik} + \tilde{M}^2_{jk} - t\,(\Pi^\mathcal{H})_{ik}\nonumber\\
 &\equiv& t\, (P^\mathcal{\tilde{H}})_{ik} + (M^2_H)_{jk}
\end{eqnarray}
where we have used eq.~(\ref{eq::goldstonetheorem}) and defined the physical Higgs mass matrix
\begin{eqnarray}
 (M^2_H)_{ij} &=&  \frac{1}{2} \lambda^\mathcal{H}_{ijkl} v_k v_l 
  - \frac{1}{6 v^2}\ \lambda^\mathcal{H}_{klmn} v_k v_l v_m v_n\ \Pi^\mathcal{H}_{ij} \,.
\end{eqnarray}
Note that $P^\mathcal{G}M^2_H=0$ and therefore the Higgs mass matrix,
as we have defined it, can have non-zero entries only on the subspace $\mathcal{R}^\mathcal{\tilde{H}}$.
Also the parameter $\mu^2_\mathcal{H}$ does not appear anymore in the definition 
of $M_H^2$, as desired.
For the fields $S_i$ no peculiarities occur, since they have no vev. 
Their mass matrix is essentially given by $\tilde{M}^2$:
\begin{eqnarray}
 (M^2_\mathcal{S})_{ij} &=&  \frac{1}{2} \lambda^\mathcal{S}_{ijkl} v_k v_l 
   - (\mu^2(\one-\Pi^\mathcal{H}))_{ij}\,.
\end{eqnarray}
The up to quadratic terms of the scalar potential can now be written as
\begin{eqnarray}
  V(\Phi) &=& t \ v_i H_i + \frac{1}{2} (M^2_H)_{ij}H_i H_j + \frac{1}{2}t \ H_i H_i 
             + \frac{1}{2}t \ G_iG_i + \frac{1}{2} (M^2_S)_{ij}S_i S_j + \mathcal{O}(\Phi^3) \
\end{eqnarray}
which is the appropriate parametrization for renormalizing the theory.

\section{Three-loop gauge $\beta$ function for the Georgi-Glashow $SU(5)$ model }\label{sec::betasu5}
For performing a consistent three-loop RGE analysis, apart from the two-loop GUT matching corrections
also the three-loop gauge $\beta$ function for the Georgi-Glashow model is needed.
The authors of ref.~\cite{Pickering:2001aq} give a general formula
for the gauge $\beta$ function of a general single gauge coupling theory.
Specifiying the group theory factors that appear there to the Georgi-Glashow 
model and insering the scalar self-couplings from eq.~(\ref{eq::su5lam}),
as well as the Yukawa coupling from eq.~(\ref{eq::su5yuk}) into their general result
gives us the desired $\beta$ function including scalar self-couplings and
Yukawa corrections:

\begin{eqnarray}\label{eq::su5gaugebeta}
\frac{1}{2} \frac{d}{d t} \frac{\alpha}{4\pi} &=& 
    -\frac{40}{3} \left(\frac{\alpha}{4\pi}\right)^{2}\  \\
  &&-\frac{1184}{15} \left(\frac{\alpha}{4\pi}\right)^{3}\ 
    +\left[-\frac{9}{2} \left(\frac{y_t}{4\pi}\right)^2 
           - 5\left(\frac{y_b}{4\pi}\right)^2  \right]\, \left(\frac{\alpha}{4\pi}\right)^{2}\nonumber\\
  &&-\frac{1007357}{1080} \left(\frac{\alpha}{4\pi}\right)^{4} \nonumber\\
     &&+\Bigg[-\frac{1323}{4} \left(\frac{y_t}{4\pi}\right)^2 
           - \frac{3617}{10}\left(\frac{y_b}{4\pi}\right)^2  \nonumber\\
           &&\hspace{0.5cm}+ \frac{155}{96}\frac{A}{(4\pi)^2} + \frac{11}{20}\frac{b}{(4\pi)^2} 
           + \frac{125}{12}\frac{B}{(4\pi)^2}  + \frac{25}{4}\frac{c}{(4\pi)^2}  
           \Bigg]\, \left(\frac{\alpha}{4\pi}\right)^{3} \nonumber\\
  &&+\Bigg[ \frac{51}{4} \left(\frac{y_t}{4\pi}\right)^4  
           +\frac{47}{4} \left(\frac{y_b}{4\pi}\right)^4 
           +\frac{839}{8} \frac{y_t^2 y_b^2}{(4\pi)^4} \nonumber\\
      &&\hspace{0.5cm}-\frac{493}{11520}\frac{A^2}{(4\pi)^4} - \frac{47}{144} \frac{A B}{(4\pi)^4}  
           -\frac{1}{12} \frac{b^2}{(4\pi)^4} - \frac{65}{36} \frac{B^2}{(4\pi)^4}
          - \frac{851}{200} \frac{c^2}{(4\pi)^4} \Bigg]\, \left(\frac{\alpha}{4\pi}\right)^{2}\,. \nonumber
\end{eqnarray}
The first line of this equation represents the one-loop result,
the second line the two-loop result and the rest corresponds to
the three-loop corrections. 
Since the Yukawa couplings enter the gauge $\beta$ function starting from two-loop level only,
it is enough to employ the one-loop RGEs for the Yukawa couplings
for the precision we are aiming at. These can be derived in a similar
manner from the general formula in ref.~\cite{Machacek:1983fi}:
\begin{eqnarray}
 \frac{d y_t}{d t} &=& y_t \Bigg[-\frac{108}{5}\left(\frac{\alpha}{4\pi}\right)
                          - 6 \left(\frac{y_b}{4\pi}\right)^2 
                          + 9 \left(\frac{y_t}{4\pi}\right)^2 \Bigg]\,, \nonumber\\[0.5cm]
 \frac{d y_b}{d t} &=& y_b \Bigg[- 18 \left(\frac{\alpha}{4\pi}\right)
                           + 11 \left(\frac{y_b}{4\pi}\right)^2 
                           - \frac{9}{2} \left(\frac{y_t}{4\pi}\right)^2  \Bigg]\,.
\end{eqnarray}
The scalar self-couplings $A,B$ and $c$ that appear in eq.~(\ref{eq::su5gaugebeta})
only at the three-loop level are approximated as constants in our analysis
by replacing them by their relations to the physical mass parameters $M_\Sigma, M_{24}, \mhc,\mx$, 
and the gauge coupling $\alpha$
by using eqs.~(\ref{eq::SU5scalarmasses}) and (\ref{eq::SU5gaugebosonmass}).
The scalar self-coupling $b$ that appears here can be approximated similarly
by a constant using the SM Higgs mass $M^2_{{\rm H,SM}}$ and the mass of the $W$ boson $M_W$:
\begin{eqnarray}
 b &=& \frac{3}{4}g^2\frac{M^2_{{\rm H,SM}}}{M_W^2} \,.
\end{eqnarray}

\end{appendix}



\end{document}